\documentclass[%
reprint,
groupedaddress,
nofootinbib,
 amsmath,amssymb,
 aps,
]{revtex4-2}

\usepackage{graphicx}% Include figure files
\usepackage{dcolumn}% Align table columns on decimal point
\usepackage{bm}% bold math
\usepackage{hyperref}% add hypertext capabilities
\usepackage[mathlines]{lineno}% Enable numbering of text and display math
\usepackage{booktabs}
\usepackage{physics}
\usepackage{rotating}

\DeclareMathAlphabet\mathbfcal{OMS}{cmsy}{b}{n}

\usepackage{xcolor}

\newcommand{\gw}{\mathrm{gw}}
\newcommand{\SFR}{\mathrm{SFR}}
\newcommand{\BBH}{\mathrm{BBH}}
\newcommand{\BNS}{\mathrm{BNS}}
\newcommand{\NSBH}{\mathrm{NSBH}}
\newcommand{\Hz}{\mathrm{Hz}}
\newcommand{\CBC}{\mathrm{CBC}}
\begin{document}

\preprint{LIGO/123-QED}

%\title{Astrophysical Gravitational-Wave Background:\\Joint Estimate of Multiple Components }
%\title{Jointly Estimating Multiple Components and Population Properties of Astrophysical Gravitational-Wave Background}%Force line breaks with \\
\title{Estimating Astrophysical Population Properties using a multi-component Stochastic Gravitational-Wave Background Search}

\author{Federico De Lillo}
 \email{federico.delillo@uclouvain.be}
\author{Jishnu Suresh}%
 \email{jishnu.suresh@uclouvain.be}
\affiliation{%
 Centre for Cosmology, Particle Physics and Phenomenology (CP3),\\
Universit\'e catholique de Louvain, Louvain-la-Neuve, B-1348, Belgium
}%

% \collaboration{CLEO Collaboration}%\noaffiliation

\date{\today}% 

\begin{abstract}
The recent start of the fourth observing run of the LIGO-Virgo-KAGRA (LVK) collaboration has reopened the hunt for gravitational-wave (GW) signals, with one compact-binary-coalescence (CBC) signal expected to be observed every few days. Among the signals that could be detected for the first time there is the stochastic gravitational-wave background (SGWB) from the superposition of unresolvable GW signals that cannot be detected individually. In fact, multiple SGWBs are likely to arise given the variety of sources, making it crucial to identify the dominant components and assess their origin. However, most search methods with ground-based detectors assume the presence of one SGWB component at a time, which could lead to biased results in estimating its spectral shape if multiple SGWBs exist. Therefore, a joint estimate of the components is necessary. In this work, we adapt such an approach and analyse the data from the first three LVK observing runs, searching for a multi-component isotropic SGWB. We do not find evidence for any SGWB and establish upper limits on the dimensionless energy parameter $\Omega_{\gw}(f)$ at 25 Hz for five different power-law spectral indices, $\alpha = 0, \, 2/3,\, 2,\, 3,\, 4$, jointly. For the spectral indices $\alpha = 2/3,\, 2, \, 4$, corresponding to astrophysical SGWBs from CBCs, r-mode instabilities in young rotating neutron stars, and magnetars, we draw further astrophysical implications by constraining the ensemble parameters $K_{\CBC}, \, K_{\rm r-modes}, \, K_{\rm magnetars}$, defined in the main text.
\end{abstract}

%In this work, we adopt a component-separation method to analyse data from the first three LVK observing runs and search for multi-component isotropic SGWB.

%In this work, we analyse the data from the first three LVK observing runs, searching for a multi-component isotropic SGWB and jointly estimating its properties.

\maketitle

%\tableofcontents

\section{Introduction}
\label{sec:introduction}

Gravitational-wave (GW) astronomy with ground-based interferometric detectors is rapidly entering its golden age. Following the first three observing runs (O1, O2, and O3 \cite{LIGO_O3_PhysRevD.102.062003, Virgo_O3_Virgo:2022ypn}) of the LIGO-Virgo-KAGRA collaboration \cite{Advanced_LIGO_LIGOScientific:2014pky, adavnced_Virgo_VIRGO:2014yos, KAGRA_detector_KAGRA:2020agh} (to which in the incoming years the approved LIGO-India \cite{LIGO_India:Saleem_2021} will join), around 90 GW signals from compact binary coalescences (CBCs) have been detected by the network and collected in the third Gravitational-Wave Transients Catalogue (GWTC-3 \cite{GWTC-3_LIGOScientific:2021djp}), together with the implications for the compact-binary populations \cite{GWTC-3_pop_KAGRA:2021duu}. In addition to that, the fourth LVK observing run (O4) recently started on 24 May 2023. This run is expected to last twenty months (including two months of commissioning the middle) and to have a CBC detection every two to three days \cite{Observing_scenario_O4_O5_Kiendrebeogo:2023hzf}. However, due to their intrinsic faintness and the limited detector sensitivity, most of the GW signals cannot be detected individually or resolved. The incoherent superposition of these unresolvable signals is expected to create a persistent stochastic gravitational-wave background (SGWB) signal, which many ongoing experiments aim to probe in a broad range of frequencies. A SGWB can be generated by large variety of phenomena of astrophysical (such as CBCs \cite{CBC_Rosado:2011kv, CBC_Zhu:2011bd, CBC_Marassi:2011si, CBC_Wu:2011ac, CBC_Zhu:2012xw}, isolated neutron stars (NSs) \cite{isolated_NSs_Rosado:2012bk, isolated_NSs_magnetars_Regimbau:2001kx, magnetars_Marassi:2010wj, magnetars_2013PhRvD..87d2002W}, NS modes \cite{r-modes_Ferrari:1998jf, r-modes_Zhu:2011pt, f-modes_Lasky:2013jfa}, core collapses to supernovae \cite{Supernovae_Buonanno:2004tp, Supernovae_2004MNRAS.351.1237H, supernovae_Sandick:2006sm, supernovae_Marassi:2009ib, Supernovae_Zhu:2010af}, stellar core collapses to black holes \cite{core_collapse_Ferrari:1998ut,core_collapse_Crocker:2015taa, core_collapse_Crocker:2017agi}) or cosmological (such as cosmic strings \cite{cosmic_strings_Kibble:1976sj, cosmic_strings_Sarangi:2002yt, cosmic_strings_Damour:2004kw, cosmic_strings_Siemens:2006yp}, first order phase transitions \cite{FOPT_Marzola:2017jzl, FOPT_VonHarling:2019rgb}, primordial black holes \cite{PBH_Mandic:2016lcn, PBH_Clesse:2018ogk,PBH_Bagui:2021dqi, PBH_Mukherjee:2021itf, PBH_Mukherjee:2021ags}, domain walls \cite{domain_walls_Martin:1996ea,domain_walls_An:2023idh}, inflation \cite{inflation_1979JETPL, inflation_Bar-Kana:1994nri, inflation_Turner:1996ck}, pre big bang models \cite{pre_big_bang_Gasperini:1992em, pre_big_bang_Mandic:2005bd, pre_big_bangGasperini:2016gre}) origin and hence exist as a superposition of different components. As a consequence, after the first SGWB detection, it will be mandatory to identify the dominant components and their origins.

Recently, several collaborations working with pulsar timing arrays (PTAs) have claimed the evidence of a SGWB signal within their data \cite{Nanograv_SGWB_NANOGrav:2023gor, EPTA_SGWB_Antoniadis:2023rey, PPTA_SGWB_Reardon:2023gzh, CPTA_SGWB_Xu:2023wog}. Yet, the data need to be more comprehensive to affirm the source of the SGWB excess under the assumption of a single component being present. A ground-based GW detector experiment will face similar issues as data pile up and a SGWB signal emerges. In such a scenario, current searches assuming the presence of a single SGWB component at a time may lead to biased measurements of its intensity if multiple components are present \cite{component_separation_Ungarelli:2001xu, component_separation_CBC_Mandic:2012pj, simultaneous_astro_cosmo_Martinovic:2020hru}. 
In light of this, the spectra would need to be estimated jointly, and component separation methods, based on Fisher matrix formalism, have already been developed for this scope in the past \cite{Joint_iso_methods_Parida:2015fma, Joint_aniso_Suresh:2021rsn, separation_LISA_Boileau:2021sni}. In this work, we consider the formalism as mentioned above and apply it to the data from the first three LVK observing runs to estimate jointly and set upper limits on the amplitudes of different SGWB, assuming them being isotropic, and their spectral shape following a power-law in frequency. Then, we use the results of the multi-component analysis to derive constraints on ensemble properties of some astrophysical sources associated with a subset of the considered SGWBs, namely the CBCs, NS r-mode instabilities, and magnetars.

The rest of the paper is organised as follows. In section \ref{sec:astro_SGWB}, we present the astrophysical sources and expressions of the SGWBs of interest. After that, in section \ref{sec:methods}, we discuss the analysis methods for a multi-component SGWB. Then, in section \ref{sec:results}, we present the results for the SGWB amplitudes and the ensemble properties, comparing them with the single-component results. In the conclusions, we summarise what we have done and present some prospects for improving this work and refining the techniques employed here. In the appendix \ref{sec:injection_study}, we include an injection study we performed to validate the methods, understand the best procedure to follow in a detection regime, and how to interpret the results in that case. 

\begin{widetext}
%JS --add a short para further motivating the search
\section{Astrophysical stochastic gravitational-wave backgrounds}
\label{sec:astro_SGWB}

The astrophysical SGWBs (AGWBs) are those backgrounds whose origin is connected to GW sources formed during the stellar history of the cosmos. Their detection and measurements may give access to properties of astrophysical populations that cannot be observed via electromagnetic astronomy. Like other SGWBs, they can be characterised by means of the dimensionless ratio of the GW energy density $\rho_{\gw}$ spectrum per logarithmic frequency unit to the critical energy needed to have a closed Universe $\rho_{c}$ \cite{Allen:1997ad}
\begin{equation}
    \label{eq:Om_gw_f_definition}
    \Omega_{\gw} (f) \equiv \frac{1}{\rho_{c}}\, \dv{\rho_{\gw}(f)}{\ln f},
\end{equation}
where $\rho_{c}=3\, H_{0}^{2}c^{2}/(8\pi\,G)$, with $H_{0}$ the Hubble parameter today, $c$ is the speed of light, and $G$ Newton's gravitational constant.

For astrophysical SGWB, $\Omega_{\gw}(f)$ can be expressed in the observer frame using Phinney's formula \cite{Phinney:2001di}, which we write here as \cite{regimbau_2011_Regimbau:2011rp, regimbau_2022_Regimbau:2022mdu}
\begin{equation}
    \label{eq:Phinney_but_Tania}
    \Omega_{\mathrm{gw}}(f) = \frac{f}{\rho_{c}\, H_{0}} \, \int_{\Theta} \, p\left(\theta\right) \dd{\theta} \int_{z_{\mathrm{min}}\left(\theta\right)}^{z_{\mathrm{max}}\left(\theta\right)} \frac{R\left(\theta, z\right)}{\left(1+z\right)E(z)}\, \dv{E_{\mathrm{gw}}(f_s, \theta)}{f_s}\eval_{f_{s} = (1+z)\, f} \dd{z},
\end{equation}
\end{widetext}
where $p(\theta)$ is the probability density function of the population parameters, $\dv*{E_{\gw}(f_{s}, \theta)}{f_{s}}$ is the GW source energy spectrum evaluated in the source frame, $R(\theta,\, z)$ is the source-frame rate per comoving volume, and the dependence on (a flat $\Lambda \mathrm{CDM}$ ignoring radiation and curvature terms) cosmology is encoded in
\begin{equation}
    \label{eq:E_z_flatLCDM_no_radiation}
    E\left(z\right) = \sqrt{\Omega_{m}\left(1+z\right)^{3} + \Omega_{\Lambda}}.
\end{equation}
The bounds in the redshift integral depend on the source parameters through the minimum and maximum emission frequencies of the source $f_{s;\,\mathrm{min}}(\theta)$ and $f_{s;\,\mathrm{max}}(\theta)$ as
\begin{align}
    \label{eq:z_min_z_max}
    &z_{\mathrm{min}}\left(\theta\right) = \max\left\{0, \frac{f_{s;\,\mathrm{min}}\left(\theta\right)}{f}-1\right\},\\
    &z_{\mathrm{max}}\left(\theta\right) = \min\left\{z_{\mathrm{max}}, \frac{f_{s;\,\mathrm{max}}\left(\theta\right)}{f}-1\right\}.
\end{align}

\begin{figure}
    \centering
    \includegraphics[width = \columnwidth]{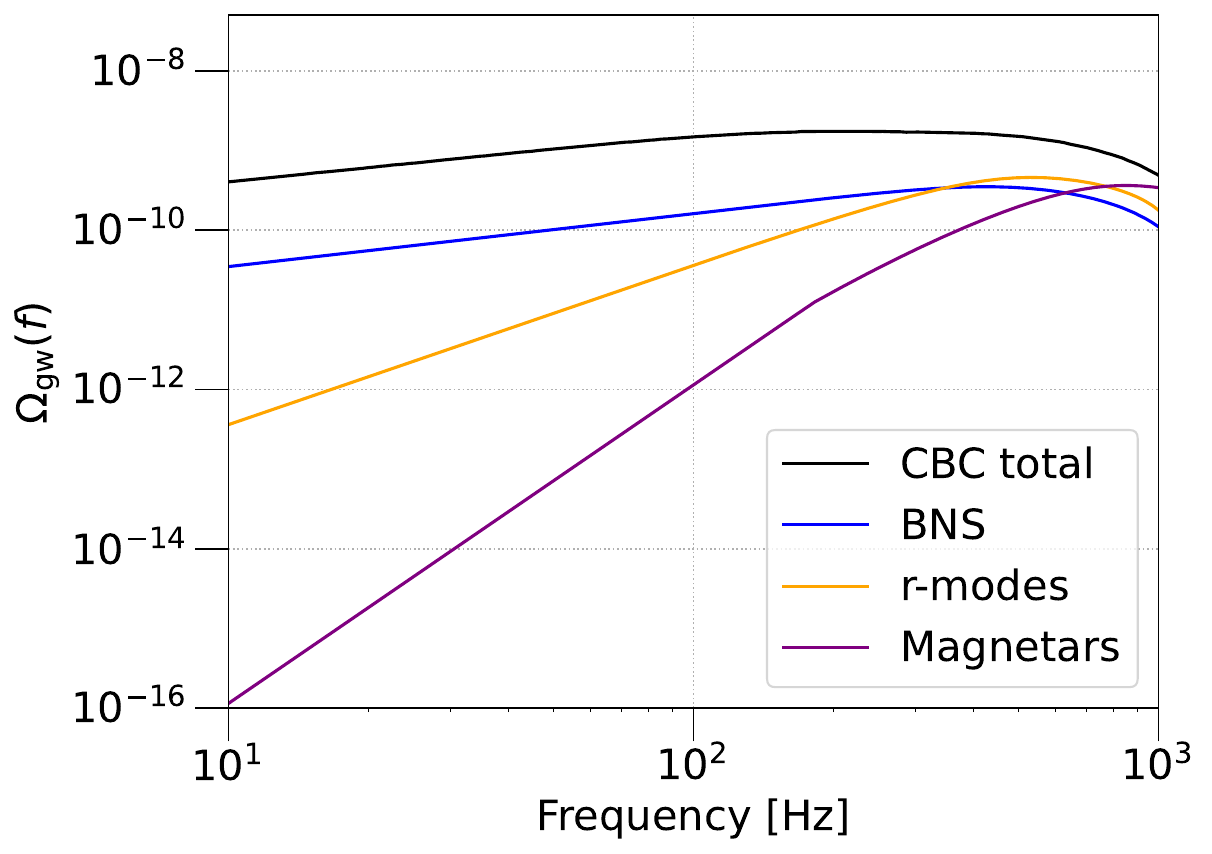}
    \caption{Landscape plot with the intensity of different astrophysical SGWBs. The black line denotes the median value of the total CBC SGWB as inferred from the GWTC-3 in \cite{GWTC-3_pop_KAGRA:2021duu}. The blue line represents the median value of the BNS SGWB, again from \cite{GWTC-3_pop_KAGRA:2021duu}. The orange line is the SGWB from r-modes from \cite{r-modes_Zhu:2011pt} with our conventions, setting $K=-5/4$ and further scaling equation \eqref{eq:r-modes_Om_gw_f} to the case where just 1\% of the young NSs enters the instability \cite{regimbau_2011_Regimbau:2011rp}. The purple line is the SGWB from magnetars, using $\varepsilon = 5\times 10^{-4}$ and $B = 10^{11}$ T, with the other parameters from \cite{Regimbau_Mandic_2008:2008nj}.} \label{fig:landscape_plot}
\end{figure}

%For the scope of this work, given that many of the distribution functions of the parameters are fledged by uncertainties, we further assume that, in the frequency range of interest, the spectrum $\Omega_{\gw}(f)$ depends only on the frequency and on the ensemble averages of the population parameters $\theta_{i}$ through a power-law functional dependence:
%\begin{equation}
%    \label{eq:Om_gw_f_ensemble_parameters}
%    \Omega_{\gw}(f) = \xi(f) \prod_{i} \expval{\theta_{i}^{c_{i}}} \approx \xi f^{\alpha} \prod_{i} \expval{\theta_{i}^{c_{i}}},
%\end{equation}
%where $\expval{\dotsc}$ denotes the ensemble average and
%\begin{equation}
%    \label{eq:xi_f}
%    \xi(f) \equiv \frac{ \Omega_{\gw}(f)}{\prod_{i} \expval{\theta_{i}^{c_{i}}}}.
%\end{equation}
For the scope of this work, given that many of the distribution functions of the parameters are fledged by uncertainties, we further assume that, in the frequency range of interest, the spectrum $\Omega_{\gw}(f)$ depends only on the frequency and on the ensemble averages of the population parameters $\theta_{i}$ of interest\footnote{The other population parameters, which we are not interested in or whose dependence cannot be written in general in the product of equation \eqref{eq:Om_gw_f_ensemble_parameters}, are assumed to be fixed and reabsorbed in the normalisation factor $\xi$.} through a power-law functional dependence:
\begin{equation}
    \label{eq:Om_gw_f_ensemble_parameters}
    \Omega_{\gw}(f) \approx \xi \left(\frac{f}{f_{\rm ref}}\right)^{\alpha} \prod_{i} \expval{\theta_{i}^{c_{i}}},
\end{equation}
where $\expval{\dotsc}$ denotes the ensemble average%\footnote{Note that, in general, $\Omega_{\gw}(f)$ may depend on population parameters $\theta_{j}$ that cannot be written as an ensemble average and do not appear in the product in equation \eqref{eq:Om_gw_f_ensemble_parameters}.}
, $f_{\rm ref}$ is a pivot frequency, and
\begin{equation}
    \label{eq:xi_f}
    \xi \equiv \frac{ \Omega_{\gw}(f_{\rm ref})}{\prod_{i} \expval{\theta_{i}^{c_{i}}}}
\end{equation}
is an overall normalisation constant.%\footnote{Note that, in general, $\xi$ may depend on population parameters $\theta_{j}$ that do not appear in the product in equation \eqref{eq:Om_gw_f_ensemble_parameters}.}
%, arising from evaluation of equation \eqref{eq:Phinney_but_Tania} when evaluated at the pivot frequency $f_{\rm ref}$.
%
The spectra of the AGWBs considered in the following are illustrated in the landscape plot in figure \ref{fig:landscape_plot}. 

\subsection{SGWB from compact binary coalescences}

The observation of binary black hole (BBH), binary neutron star (BNS), and binary neutron star black hole (NSBH) merger signals during the LVK first three observing runs has allowed to draw more precise information about the population properties of these objects \cite{GWTC-3_LIGOScientific:2021djp, GWTC-3_pop_KAGRA:2021duu}. The results suggest that the SGWB from CBCs is the dominant one in the ground-based GW detector network's most sensitive frequency band, predicting the energy densities at the reference frequency $f_{\rm{ref}}=25\, \Hz$ being $\Omega_{\BNS}(25\, \Hz)=0.6^{+1.7}_{-0.5}\times 10^{-10}$, $\Omega_{\NSBH}(25\, \Hz)= 0.9^{+2.2}_{-0.7}\times 10^{-10}$, and $\Omega_{\BBH}(25\, \Hz)=5.0^{+1.4}_{-1.8}\times 10^{-10}$ \cite{GWTC-3_pop_KAGRA:2021duu}. The SGWB from these objects can be modelled using equation \eqref{eq:Phinney_but_Tania} by explicitly expressing the energy spectrum of the sources and the merger rate.

It is a standard result that, in the quasi-circular orbit, Newtonian approximation, the energy spectrum of an inspiralling binary is given by
\begin{equation}
    \label{eq:energy_spectrum_CBC_newtonian}
    \dv{E_{\gw}}{f}(f) = \frac{\pi^{2/3}}{3G} \left(G\mathcal{M}_{c}\right)^{5/3} f^{-1/3}.
\end{equation}
However, to make predictions about the SGWB from BBHs, it is necessary to consider the full inspiral-merger-ringdown contribution to the energy spectrum, which would otherwise be underestimated. In this case, the above equation is no longer valid, and recent studies \cite{Lehoucq:2023zlt} for SGWB from BBHs usually make use of the phenomenological waveform at second post-Newtonian order from \cite{Ajith:2007kx, Ajith:2009bn} in the limit of non-spinning black holes.

The merger rates for CBCs are intrinsically related to the cosmic stellar formation rate (SFR) \cite{SFR_Vangioni:2014axa} through the convolution integral over time delays
\begin{equation}
    \label{eq:rates_CBC}
    R(z) = \int \dd{t_d} R_{f}(z_{f}(z, t_{d}))\, P_{t_{d}}(t_{d}),
\end{equation}
where the dependence on the SFR (usually in $M_{\odot}\, \mathrm{yr^{-1}\, Mpc^{-3}}$ units) is implicitly in the formation rate
\begin{equation}
    \label{eq:CBC_formation_rate}
    R_{f}(z_{f}) = r_{0} \frac{\SFR(z_{f})}{\SFR(0)}
\end{equation}
normalised to the local rate $r_{0}$ (usually in $\mathrm{yr^{-1}\, Gpc^{-3}}$ units), which can be inferred from observations. The most recent results from GWTC-3 \cite{GWTC-3_pop_KAGRA:2021duu} suggest that the BNS local merger rate is in the range 10-1700 $\mathrm{yr^{-1}\, Gpc^{-3}}$, the NSBH one between 7.8 and 140 $\mathrm{yr^{-1}\, Gpc^{-3}}$, and the BBH one, evaluated at the fiducial redshift $z=0.2$, to be in the 17.9 and 44 $\mathrm{yr^{-1}\, Gpc^{-3}}$ interval.

The integral over the time-delays distribution is necessary to account for the time elapsing between the formation of a stellar binary system and its evolution towards a compact binary system. The time-delay distribution is usually assumed to follow a power-law $p_{t_{d}}\propto t_{d}^{-1}$ \cite{time_delay_Chruslinska:2017odi}, with the maximum time delay equal to the Hubble time, and the minimum time delay for BNS and NSBH being 20 Myr, and 50 Myr for BBH.
In the case of BBH, the merger rate integral may also be further weighted by including a metallicity cut for stars forming at $Z<0.1\, Z_{\odot}$ \cite{metalllicity_Mapelli:2019bnp, metallicity_Chruslinska:2018hrb}. These aspects and parameters still need to be completely understood and are usually inferred from population syntheses \cite{population_synthesis_Santoliquido:2020axb} in combination with
electromagnetic observations. A recent work \cite{Lehoucq:2023zlt} considers most of the uncertainties in these parameters and carefully evaluates their impact when predicting the spectrum of a SGWB from BBH and BNS in the ground-based detectors frequency range.

The parameter space in equation \eqref{eq:Phinney_but_Tania} for a compact binary usually ranges between 15 (BBH) and 17 (BNS, NSBH) parameters. Current studies for SGWB from CBCs usually limit to average the rates and the energy spectrum over the two components masses $m_{1}$ and $m_{2}$ of the binary (in the BBH case, it is possible to include spin as well through the above mentioned phenomenological waveform, but it is usually considered to be zero, spin corrections being small \cite{CBC_Marassi:2011si, CBC_Zhu:2011bd}). Population studies from GWTC-3 \cite{GWTC-3_pop_KAGRA:2021duu} suggest a preference for a power-law-plus-peak model but do not exclude a broken power-law model for BBH components mass function. The mass function for BNS components is expected instead to exhibit broader features, which for SGWB studies can be taken to be uniformly distributed between 1 and 2.5 $M_{\odot}$. For NSBH, GWTC-3 studies have used the same mass function for the NS component and a logarithmically uniform distribution of black hole masses between 5 and 50 $M_{\odot}$.

\begin{widetext}
For this work, when constraining the population properties of a CBC background, we will use a frequency range where this background is well described by the inspiral phase, assuming hence the functional dependence 
%\begin{equation}
%    \label{eq:Om_CBC_pop_par}
%    \Omega_{\gw,\, j}(f) = \xi_{j}\, f^{2/3}\, r_{0, \, j}\expval{\mathcal{M}_{c}^{5/3}}_{j} \equiv \xi_{j}\, f^{2/3}\, K_{j}, \qquad j=\BBH,\, \BNS,\, \NSBH,
%\end{equation}
\begin{equation}
    \label{eq:Om_CBC_pop_par}
    \Omega_{\gw,\, j}(f) = \xi_{j}\, \left(\frac{f}{f_{\rm ref}}\right)^{2/3} r_{0, \, j}\expval{\mathcal{M}_{c}^{5/3}}_{j} \equiv \xi_{j}\, \left(\frac{f}{f_{\rm ref}}\right)^{2/3} K_{j}, \qquad j=\BBH,\, \BNS,\, \NSBH,
\end{equation}
where the product $K_{j} \equiv r_{0, \, j}\expval{\mathcal{M}_{c}^{5/3}}_{j}$ is assumed to be a free parameter to be constrained. More specifically, we will focus on the whole population of CBCs and constrain an effective $K_{\CBC} \equiv \sum_{j} K_{j}$. 
\end{widetext}

\subsection{Magnetars}

Magnetars are neutron stars that are formed with a very intense magnetic field (of the order of $10^{10}-10^{11}$ T). They were first proposed in \cite{magnetars_proposal_1992ApJ} as a candidate to explain soft gamma repeaters and anomalous x-ray pulsars. The list of known magnetars is collected in the McGill magnetars catalogue \cite{magnetar_catalogue_Olausen:2013bpa}.
\begin{widetext}
The intense magnetic field is expected to induce a quadrupolar deformation in the rapidly spinning neutron star, which in turn decelerates through magnetic dipole torque and GW production, with energy spectrum \cite{Regimbau_Mandic_2008:2008nj, magnetars_Marassi:2010wj, magnetars_2013PhRvD..87d2002W}
\begin{equation}
\label{eq:magnetar_energy_spectrum}
\dv{E_{\gw}}{f} = I \pi^2 f^3 \, \left( \frac
{5c^2 R^6}{192 \pi^2 G I^2} \frac{4\pi B^2}{\mu_0 \varepsilon^2}\, \sin^{2}{\alpha} +f^2\right)^{-1},\quad f \in \left[0-\frac{2}{P_{0}}\right],
\end{equation}
where $I$ is the magnetar moment of inertia around the rotation axis, $R$ is the magnetar radius, $B$ the (poloidal) magnetic field of the star, $\varepsilon$ the dimensionless ellipticity, quantifying the deviation from spherical symmetry, $\alpha$ the ``wobble angle'' between the magnetar spin and magnetic axes, $P_{0}$ the initial rotation period of the magnetar, and $\mu_{0}$ the vacuum magnetic permeability. Recurrent reference values in the literature for some of the magnetar parameters are $I = 1 \times 10^{38}\, {\rm kg\, m^{2}}$, $R=10\, {\rm km}$, $\alpha = \pi/2$, and $P_{0}= 1\, {\rm ms}$ \cite{Regimbau_Mandic_2008:2008nj, magnetars_Marassi:2010wj, magnetars_2013PhRvD..87d2002W}.
\end{widetext}
The first term in the brackets comes from the rotational energy loss due to electromagnetic dipole radiation, while the second term is due to GW emission. The ellipticity may further depend on the magnetic field in different ways, based on whether the magnetic field configuration is expected to be poloidal dominated \cite{poloidal_Bonazzola:1995rb}, a twisted-torus one \cite{twisted_torus_Braithwaite:2004ubj}, or toroidal dominated \cite{toroidal_Stella:2005yz}.

When deriving the expression for the SGWB from magnetars from equation \eqref{eq:Phinney_but_Tania}, given that the GW source starts emitting after its birth, it is possible to write the cosmic rate $R(\theta, z)$ as $R_{\rm magnetar}(\theta, z) = \lambda_{\rm magnetar}\, \SFR(z)$, where $\lambda_{\rm magnetar}$ is the fraction per solar mass of the progenitor mass that is converted in magnetars. This quantity can also be written as a function of the mass of the NS progenitors $\lambda_{\rm NS}$ and the fraction of NSs born as magnetars $f_{\rm magnetars}$, namely $\lambda_{\rm magnetar}= f_{\rm magnetars}\, \lambda_{\rm NS}$.

\begin{widetext}
The resulting expression of $\Omega_{\gw}(f)$ in the case where the dominant rotational energy loss mechanism is the dipole magnetic torque is then \cite{Regimbau_Mandic_2008:2008nj, magnetars_2013PhRvD..87d2002W}
\begin{equation}
    \label{eq:om_gw_magnetars}
    \Omega_{\gw}(f) = \frac{\mathcal{K}}{\rho_{c}\, H_{0}}\,f^{4} \int_{z_{\min}}^{z_{\max}} \frac{\lambda_{\rm magnetars}\, \SFR(z)\, (1+z)^{2}}{E(z)},
\end{equation}
where
\begin{equation}
    \label{eq.K_magnetars}
    \mathcal{K} = \frac{192 \pi^{4} G}{5c^{2}}\, \frac{\mu_{0}}{4\pi}\, \expval{\frac{1}{R^{6}}}\expval{I^{3}}\, \expval{\varepsilon^{2}} \expval{\frac{1}{B^{2}}} \expval{\frac{1}{\sin^{2}{\alpha}}}.
\end{equation}
In the following, we assume the following functional dependence
%\begin{equation}
%    \label{eq:magnetars_pop_par}
%    \Omega_{\gw, \, \mathrm{magnetars}}(f) = \xi_{\rm magnetars}\, f^{4}\, \expval{\varepsilon^{2}}\, \expval{\frac{1}{B^{2}}} \equiv  \xi_{\rm magnetars}\, f^{4}\, K_{\rm magnetars},
%\end{equation}
\begin{equation}
    \label{eq:magnetars_pop_par}
    \Omega_{\gw, \, \mathrm{magnetars}}(f) = \xi_{\rm magnetars}\, \left(\frac{f}{f_{\rm ref}}\right)
^{4} \expval{\varepsilon^{2}} \expval{\frac{1}{B^{2}}} \equiv  \xi_{\rm magnetars}\, \left(\frac{f}{f_{\rm ref}}\right)
^{4} K_{\rm magnetars},
\end{equation}
where $K_{\rm magnetars} \equiv \sqrt{\expval{\varepsilon^{2}}\, \expval{1/B^{2}}}$ is assumed to be a free parameter to be constrained.
\end{widetext}

\subsection{r-mode instabilities}

In the pioneering works \cite{r-modes_Andersson:1997xt, r-modes_Friedman:1997uh}, it was discovered that the emission of gravitational radiation induces instability in the r-modes of young rapidly rotating NSs. The first studies about the GW emission (and SGWB) related to the r-mode instability were performed in \cite{r-modes_Owen:1998xg, r-modes_Ferrari:1998jf}, showing that the GW emission can be modelled by two parameters only, namely the angular velocity of the NS $\Omega$ and the parameter $\alpha$, related to the r-modes amplitude $h(t)$ from the $l=m=2$ current multipole $S_{22}$ dominating the GW emission. After an initial (500-s \cite{r-modes_Owen:1998xg}) phase where $\Omega$ is roughly constant, and $\alpha$ exponentially grows, the system enters a non-linear hydrodynamic regime, with $\alpha$ reaching a saturation value, that lasts around one year and eventually radiates approximately two thirds of the NS rotational energy in GWs, ceasing the r-mode instability \cite{r-modes_Owen:1998xg}. The superposition of the GW signals from young NSs during the year-long non-linear phase can give rise to a continuous SGWB \cite{r-modes_Ferrari:1998jf, regimbau_2011_Regimbau:2011rp, r-modes_Zhu:2011pt}.

The expression for $\Omega_{\gw}(f)$ is analogous to the one for SGWB from magnetars, with the main difference within the fraction of initial mass progenitors converted into neutron stars $\lambda_{\rm NS}=9\times 10^{-3}\, M_{\odot}^{-1}$ \cite{regimbau_2011_Regimbau:2011rp}, and the energy spectrum \cite{r-modes_Owen:1998xg}
\begin{equation}
    \label{eq:r-modes_energy_spectrum}
    \dv{E_{\gw}}{f}\eval_{\rm r-modes} \approx \frac{4}{3}\, \frac{f}{f_{\max}^{2}}\, E_{K},
\end{equation}
where $E_{K}$ is the rotational kinetic energy of the NS assuming a Keplerian angular velocity $\Omega_{K}=2\sqrt{\pi G \bar{\rho}_{\rm NS}}/3$ (namely the angular velocity at which the star starts shedding mass at the equator), and $f_{\max}$ the corresponding maximal GW frequency\footnote{Note that $E_{K}\propto \Omega_{K}^{2} \propto f_{\max}^{2}$, and hence equation \eqref{eq:r-modes_energy_spectrum} does not depend on $f_{\max}$.}.

\begin{widetext}
By using the results from more recent studies and simulations about gravitational radiation from r-mode instability \cite{r-modes_Sa_Tome:2006hn}, which allow for saturation values of $\alpha$ being less than unity, and adapting the resulting expression for $\Omega_{\gw}(f)$ derived in \cite{r-modes_Zhu:2011pt} to our notation and convention, it is possible to express the spectrum as (assuming all NSs having a mass equal to $1.4 M_{\odot}$, a radius $R=12.53$ km, and a polytropic equation of state $p=k\rho^{2}$ \cite{r-modes_Owen:1998xg, r-modes_Ferrari:1998jf, r-modes_Sa_Tome:2006hn, r-modes_Zhu:2011pt})
\begin{equation}
    \label{eq:r-modes_Om_gw_f}
    \Omega_{\gw}(f)\eval_{\rm r-modes} = \frac{16\pi^{3} c}{3H_{0}^{3}}\times 96.7 \times \, \expval{\frac{1}{K+2}}\, f^{2}\,\int_{z_{\min}}^{z_{\max}}\dd{z} \frac{\SFR(z)\,\lambda_{\rm NS}}{E(z)}
\end{equation}
where $(K+2)^{-1} \propto \alpha^{2}$ when it saturates (see \cite{r-modes_Sa_Tome:2006hn} for more details about $K$ definition), with $-5/4 \leq K \ll 10^{13}$. The minimum value $K=-5/4$ corresponds to the smallest amount of differential rotation at the time when the r-mode instability is created, while the upper bound $K\ll 10^{-13}$ encodes the condition that the r-mode angular momentum is much smaller than the angular momentum of the unperturbed star \cite{r-modes_Sa_Tome:2006hn}.
In the following, when performing the analysis, we consider the following expression for the spectrum
%\begin{equation}
%    \label{eq:r-modes_pop_par}
%    \Omega_{\gw,\, {\rm r-modes}}(f) = \xi_{\rm r-modes}\, f^{2}\,\expval{\frac{1}{K+2}} \equiv \xi_{\rm r-modes}\, f^{2}\,K_{\rm r-modes}, 
%\end{equation}
\begin{equation}
    \label{eq:r-modes_pop_par}
    \Omega_{\gw,\, {\rm r-modes}}(f) = \xi_{\rm r-modes}\, \left(\frac{f}{f_{\rm ref}}\right)
^{2}\,\expval{\frac{1}{K+2}} \equiv \xi_{\rm r-modes}\, \left(\frac{f}{f_{\rm ref}}\right)
^{2}\,K_{\rm r-modes}, 
\end{equation}
where the parameter $K_{\rm r-modes}\equiv \expval{(K+2)^{-1}}$ is the free parameter to be constrained.
\end{widetext}

\section{\label{sec:methods} Analysis methods}

We perform a search for a Gaussian\footnote{The CBC SGWB is not expected to be Gaussian in the frequency range of interest; however, this does not introduce additional biases in the analysis \cite{non-Gaussian_Drasco:2002yd}.}, stationary, unpolarised and isotropic SGWB, assuming the presence of multiple components $\{\Omega_{\alpha}(f)\}$ following a power law in frequency, such that
\begin{equation}
    \label{eq:Omega_multi_components}
    \Omega_{\gw}(f) = \sum_{\alpha} \Omega_{\alpha} w_{\alpha}(f), \quad w_{\alpha}(f) \equiv \frac{\Omega_{\alpha}(f)}{\Omega_{\alpha}(f_{\rm ref})}= \left(\frac{f}{f_{\rm ref}}\right)^{\alpha},
\end{equation}
where every component is characterised by an amplitude $\Omega_{\alpha}\equiv \Omega_{\alpha}(f=f_{\rm ref})$, with $f_{\rm ref}$ an arbitrary reference frequency, chosen to be $f_{\rm ref} = 25$ Hz in the following.
We make use of the publicly available \cite{gwosc:LIGOScientific:2019lzm,gwosc:KAGRA:2023pio} time-series data from the first three observing runs (O1, O2, and O3) of the Advanced LIGO-Hanford (H) and LIGO-Livingston (L) detectors and the Advanced Virgo (V) detector. In the same fashion as in \cite{o3_iso_KAGRA:2021kbb, O3_iso_data}, we apply both time- and frequency-domain cuts and then perform the cross-correlation search employing the publicly available algorithm in MATLAB \cite{stochasticM}.

For every detector pair $IJ$ ($I, J =$ H, L, V), called ``baseline'', we divide the time-series output in segments $s_{I}(t)$, labelled by $t$, of duration $T$, and we take their short-time Fourier transform $\tilde{s}_I(t;\,f)$, obtaining a segment-dependent cross-correlation statistic spectrum $C_{IJ}(t;f) \equiv \tilde{s}_{I}^{*}(t;\,f)\tilde{s}_{J}(t;\,f)$. In the absence of correlated noise, the expectation value of $C_{IJ}(t;f)$ over the segment can be written as a linear convolution equation
\begin{equation}
    \label{eq:CSD_as_convolution_equation}
    \expval{C_{IJ}}(t;f) = \sum_{\alpha} K_{\alpha}(t; f)\, \Omega_{\alpha}, \qquad \mathrm{or} \qquad  \mathbf{C} = \mathbf{K}\cdot \mathbf{\Omega},
\end{equation}
where
\begin{equation}
    \label{eq:kernel}
    \mathbf{K} \equiv K_{\alpha}(t; f) := \frac{T}{2} S_{0}(f) \gamma_{IJ}(f) w_{\alpha}(f),
\end{equation}
with $S_{0}(f) = (3H_0^2)/(10 \pi^2 f^3)$, and $\gamma_{IJ}(f)$ the normalised overlap reduction function~\citep{Allen:1997ad,Christensen_ORF_PhysRevD.46.5250,Flanagan_ORF_PhysRevD.48.2389} quantifying the reduction in sensitivity due to the geometry of the baseline $IJ$ and its response to the GW signal.

\begin{figure}
    \centering
    \includegraphics[width=\columnwidth]{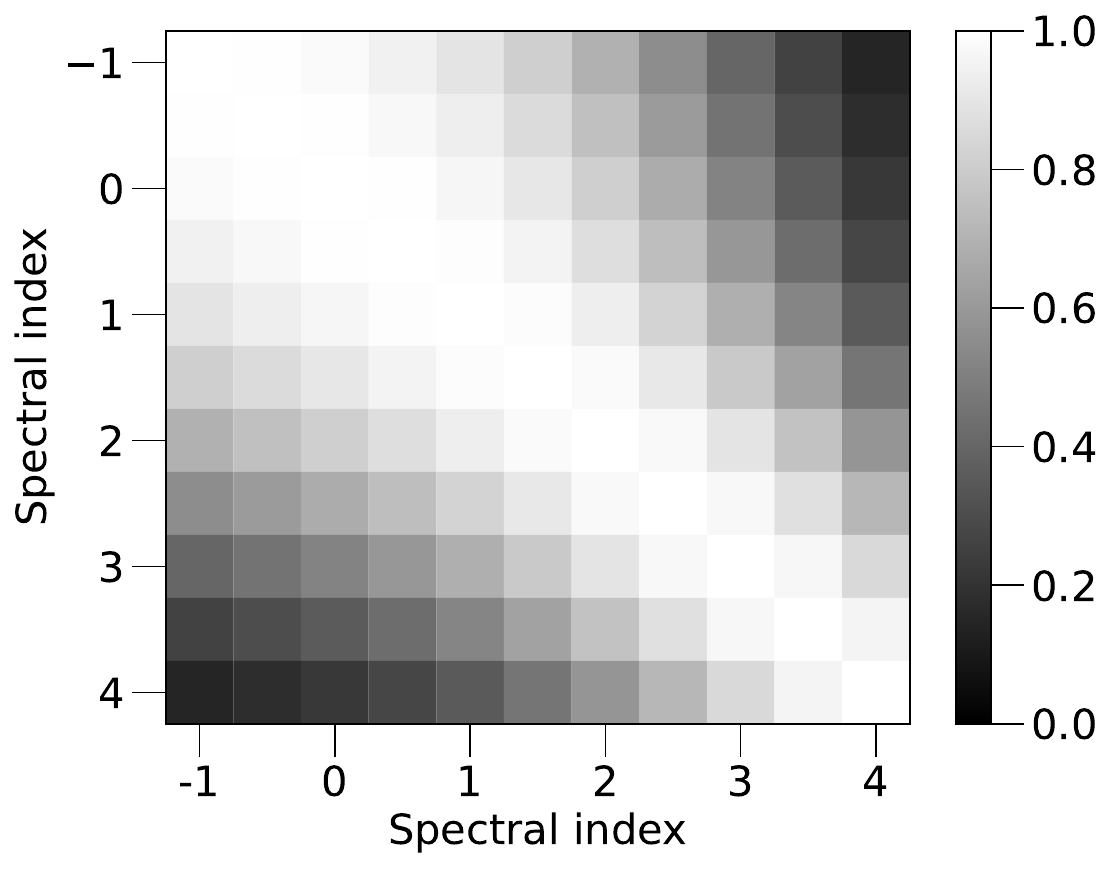}
    \caption{Preconditioned Fisher matrix for data from the first three LVK observing runs, showing the couplings among different spectral indices. The diagonal is unity by construction, with the off-diagonal element being smaller than one and positive.}
    \label{fig:coupling_matrix}
\end{figure}

The estimator for $\mathbf{\Omega}$ can be obtained as Maximum-Likelihood solution for the convolution equation \eqref{eq:CSD_as_convolution_equation}, namely \cite{Joint_iso_methods_Parida:2015fma}
\begin{equation}
    \label{eq:ML_Omega_hat}
    \hat{\mathbf{\Omega}} = \mathbf{\Gamma}^{-1}\cdot \mathbf{X},
\end{equation}
where
\begin{align}
    \label{eq:dirty_map_and_fisher}
    &X_{\alpha} = 4 \Delta f\, \sum_{I>J}\sum_{f,\,t} \frac{\gamma_{IJ}(f)\, S_{0}(f)\,\tilde{s}_{I}^{*}(t;\,f)\tilde{s}_{J}(t;\,f)}{P_{I}(t; f)\, P_{J}(t; f)}\, w_{\alpha}(f), \nonumber \\
    &\Gamma_{\alpha \alpha'} = 2 T \Delta f\,  \sum_{I>J}\sum_{f,\,t} \frac{\gamma_{IJ}^{2}(f)\, S^{2}_{0}(f)}{P_{I}(t; f)\, P_{J}(t; f)} \,w_{\alpha}(f)\,w_{\alpha'}(f),
\end{align}
where $P_{I}(t; f)$ and $P_{J}(t; f)$ are the one-sided power spectral densities of the noise in the detectors. The inversion of $\Gamma_{\alpha \alpha'}$ may lead to numerical errors. This can be avoided by preconditioning the matrix as
\begin{equation}
    \label{eq:fisher_conditioned}
    \mathbf{\Gamma} \equiv \Gamma'_{\alpha \alpha'} =
    \frac{\Gamma_{\alpha \alpha'}}{\sqrt{\Gamma_{\alpha} \Gamma_{\alpha'}}},
\end{equation}
where $\Gamma_{\alpha}\equiv \mathrm{diag}\left(\Gamma_{\alpha \alpha'}\right)$. This new coupling matrix can quantify the correlation between different models, as illustrated in figure \ref{fig:coupling_matrix}. 
Then the estimator for $\Omega_{\alpha}$ becomes
\begin{equation}
    \label{eq:Om_a_estimator_preconditioned}
    \hat{\Omega}_{\alpha} = \sum_{\alpha'} \Gamma'^{-1}_{\alpha \alpha'}\, \frac{X^{'}_{\alpha'}}{\sqrt{\Gamma_{\alpha}}},
\end{equation}
where $X'_{\alpha} = X_{\alpha}/\sqrt{\Gamma_{\alpha}}$. The covariance matrix and the standard deviation of $\hat{\Omega}_{\alpha}$ are then
\begin{align}
    \label{eq:covariance_and_variance}
    &\Sigma_{\alpha \alpha'} = \frac{\left(\Gamma_{\alpha\alpha'}'\right)^{-1}}{\sqrt{\Gamma_{\alpha}\,\Gamma_{\alpha'}}},\\
&\sigma_{\alpha} = \left[\sqrt{\mathrm{diag}(\Sigma_{\beta \beta'})}\right]_{\alpha}.
\end{align}

\begin{widetext}   
The likelihood function associated with these estimators can be assumed to be a multivariate Gaussian \cite{Joint_iso_methods_Parida:2015fma}:
\begin{equation}
    \label{eq:likelihood_Omega_joint}
    \mathcal{L} \left(\hat{\Omega}_{\alpha} | \Omega_{\alpha}\right) = 
    \frac{1}{\left(2\pi\right)^{N/2} \left(\det\Sigma \right)^{1/2}} \,
    \exp \left[-\frac{1}{2}\left(\hat{\Omega}_{\alpha}-\Omega_{\alpha}\right) \Sigma_{\alpha\alpha'}^{-1}\left(\hat{\Omega}_{\alpha'}-\Omega_{\alpha'}\right)\right].
\end{equation}
We use the above likelihood function to perform parameter estimation or set bounds on the parameters from the different considered models.
\end{widetext}
The same likelihood function can also be used to estimate the ensemble properties by using equation \eqref{eq:Om_gw_f_ensemble_parameters} and to rewrite the $\Omega_{\alpha}$ model as
%\begin{equation}
%    \label{eq:Om_a_ensemble_properties_PE}
%    \Omega_{\alpha} = \xi_{f_{\rm ref}}\, \left(\frac{f}{f_{\rm ref}}\right)^{\alpha} \prod_{i} \expval{\theta_{i}^{c_{i}}}.
%\end{equation}
\begin{equation}
    \label{eq:Om_a_ensemble_properties_PE}
    \Omega_{\alpha} = \xi_{\alpha} \prod_{i} \expval{\theta_{i}^{c_{i}}}.
\end{equation}
We have tested the method presented in this section by performing multiple sets of injections in O3 data in the frequency domain, quantifying possible sources of bias; see appendix \ref{sec:injection_study} for more details about the most significant ones.

\section{Results and Discussions}
\label{sec:results}

We present the results of the multi-component analysis for the set of spectral indices $\va*{\alpha}=\{0,\,  2/3,\, 2,\, 3,\, 4\}$ and their implications for the ensemble properties of CBCs (in the inspiral phase), pulsar r-mode instabilities, and magnetars.

\subsection{Power law energy density spectrum}%$\Omega_{\gw}$}

We have performed a search for a Gaussian, stationary, unpolarised, isotropic, multi-component SGWB following the methods presented in the previous section \ref{sec:methods} and applied them to five spectral indices, namely $\alpha = 0, \, 2/3,\, 2,\, 3,\, 4$. The indices $\alpha = 2/3,\, 2, 4$ are associated with the astrophysical SGWBs produced by CBCs \cite{CBC_Wu:2011ac, CBC_Zhu:2011bd, CBC_Rosado:2011kv, CBC_Marassi:2011si, CBC_Zhu:2012xw}, r-mode instabilities in NSs \cite{r-modes_Owen:1998xg, r-modes_Ferrari:1998jf, r-modes_Zhu:2011pt}, and magnetars \cite{isolated_NSs_magnetars_Regimbau:2001kx, magnetars_Marassi:2010wj, magnetars_2013PhRvD..87d2002W}, respectively, as described in section \ref{sec:astro_SGWB}. The remaining $\alpha = 0,\, 3$ can be associated with other SGWBs with different origins ($\alpha = 0$ to inflationary \cite{inflation_1979JETPL, inflation_Bar-Kana:1994nri, inflation_Turner:1996ck} or cosmic strings \cite{cosmic_strings_Kibble:1976sj, cosmic_strings_Sarangi:2002yt, cosmic_strings_Damour:2004kw, cosmic_strings_Siemens:2006yp} SGWBs, $\alpha =3$ to core-collapse to supernova \cite{supernovae_Marassi:2009ib, supernovae_Sandick:2006sm, Supernovae_Zhu:2010af, Supernovae_Buonanno:2004tp, Supernovae_2004MNRAS.351.1237H} or pre-big bang models \cite{pre_big_bang_Gasperini:1992em, pre_big_bang_Mandic:2005bd, pre_big_bangGasperini:2016gre}), for which we do not draw any additional implication. Moreover, the SGWBs associated with these two spectral indices are usually constrained by searches using the ground-based detectors \cite{o3_iso_KAGRA:2021kbb}, and it may be interesting to compare the standard results with those of the multi-component analysis method. For the reasons presented in the appendix \ref{sec:injection_study} in light of the implications for astrophysical SGWBs, we have restricted the frequency range of the search to $20-100$ Hz, where the astrophysical signal $\Omega_{\gw}(f)$ can be approximated by a power law\footnote{Note that, based on the discussion in \cite{o3_iso_KAGRA:2021kbb}, the choice of this frequency band may limit the search sensitivity for $\alpha > 2/3$, resulting in the upper limits being more conservative.}.

The estimators from analysing every possible combination of spectral indices can be read in table \ref{tab:PL_analysis_estimators_table}. The first five lines can be interpreted as results from the single-component analysis. The small difference in the $\alpha = 0,\,2/3,$ and $3$ cases from \cite{o3_iso_KAGRA:2021kbb} is only to be attributed to the frequency range adopted for our analysis being $20-100$ Hz (when we make use of the $20-1726$ Hz range used in \cite{o3_iso_KAGRA:2021kbb}, we recover the same estimators). Compared to the single-component case, the uncertainty in the estimate of the components of the SGWB is, in general, larger in the multi-component analysis, leading to more conservative estimates but still compatible with the single-component ones. Another factor influencing the magnitude of the estimator uncertainties when different combinations of components (with the number of components fixed) are considered is the distance of the spectral indices in the power-law space. As an example of this effect, consider the $\hat{\Omega}_{0}$ uncertainty in the four combinations $\{\alpha = 0,\,2/3\}$, $\{\alpha = 0,\,2\}$, $\{\alpha = 0,\,3\}$ and $\{\alpha = 0,\,4\}$, with the second spectral index getting more and ``distant'' from $\alpha = 0$ and the uncertainty getting lower and lower.

Consistently with previous analyses involving data from the first three LVK observing runs, there is no evidence for a signal in any of the examined combinations of the spectral indices. We have verified this for every combination by performing a Bayesian model comparison between noise-only and signal hypotheses, with the Bayes factor always favouring noise over signal. As a consequence, we have set $95\%$-confidence Bayesian upper limits $\Omega_{\alpha}^{95\%}$ for every component for every combination, using a uniform prior between $0$ and $10^{-6.5}$ for each of them.
The upper limits are summarised in table \ref{tab:unified_UL_table}. Noticeably, the upper limits become more stringent when including more components in the analysis, following the opposite trend compared to that of the estimator uncertainties. This seems to contradict the naive picture that considering more components should decrease the constraining power of the search. On the other hand, this could be argued by claiming that the power splits among the different components, leading to lower upper limits when more components are included in the analysis, as previously observed in \cite{GR_extra_pol__LIGOScientific:2018czr}. 
Alternatively, this trend could also arise from this particular data realisation being noise-dominated \cite{Joint_aniso_Suresh:2021rsn}. Still, in the presence of a signal (or excesses) in the data, the multi-component analysis provides unbiased estimators and correct parameter estimation results (and hence upper limits), in contrast to the biased single-component ones, see appendix \ref{sec:injection_study}.

\begin{table*}
\scriptsize
\makebox[\textwidth][c]{
\begin{tabular}{llllll}
\toprule
{} &              $\hat{\Omega}_{0}$ &            $\hat{\Omega}_{2/3}$ &              $\hat{\Omega}_{2}$ &               $\hat{\Omega}_{3}$ &                  $\hat{\Omega}_{4}$ \\
\midrule
$\alpha=\{0\}$               &    $(1.5 \pm7.5)\times 10^{-9}$ &                               - &                               - &                                - &                                   - \\
$\alpha=\{2/3\}$             &                               - &  $(2.3 \pm56.2)\times 10^{-10}$ &                               - &                                - &                                   - \\
$\alpha=\{2\}$               &                               - &                               - &   $(-1.3 \pm2.5)\times 10^{-9}$ &                                - &                                   - \\
$\alpha=\{3\}$               &                               - &                               - &                               - &  $(-9.8 \pm10.3)\times 10^{-10}$ &                                   - \\
$\alpha=\{4\}$               &                               - &                               - &                               - &                                - &      $(-4.0 \pm3.4)\times 10^{-10}$ \\
\hline
$\alpha=\{0, 2/3\}$          &    $(4.4 \pm4.6)\times 10^{-8}$ &   $(-3.2 \pm3.4)\times 10^{-8}$ &                               - &                                - &                                   - \\
$\alpha=\{0, 2\}$            &    $(1.6 \pm1.4)\times 10^{-8}$ &                               - &   $(-5.8 \pm4.6)\times 10^{-9}$ &                                - &                                   - \\
$\alpha=\{0, 3\}$            &    $(9.5 \pm9.5)\times 10^{-9}$ &                               - &                               - &    $(-1.8 \pm1.3)\times 10^{-9}$ &                                   - \\
$\alpha=\{0, 4\}$            &    $(6.1 \pm8.2)\times 10^{-9}$ &                               - &                               - &                                - &      $(-5.1 \pm3.7)\times 10^{-10}$ \\
$\alpha=\{2/3, 2\}$          &                               - &    $(1.7 \pm1.4)\times 10^{-8}$ &   $(-8.3 \pm6.1)\times 10^{-9}$ &                                - &                                   - \\
$\alpha=\{2/3, 3\}$          &                               - &    $(8.4 \pm8.1)\times 10^{-9}$ &                               - &    $(-2.1 \pm1.5)\times 10^{-9}$ &                                   - \\
$\alpha=\{2/3, 4\}$          &                               - &    $(5.0 \pm6.6)\times 10^{-9}$ &                               - &                                - &      $(-5.6 \pm4.0)\times 10^{-10}$ \\
$\alpha=\{2, 3\}$            &                               - &                               - &    $(7.6 \pm7.2)\times 10^{-9}$ &    $(-3.9 \pm2.9)\times 10^{-9}$ &                                   - \\
$\alpha=\{2, 4\}$            &                               - &                               - &    $(3.1 \pm4.3)\times 10^{-9}$ &                                - &      $(-7.4 \pm5.8)\times 10^{-10}$ \\
$\alpha=\{3, 4\}$            &                               - &                               - &                               - &     $(2.4 \pm3.7)\times 10^{-9}$ &       $(-1.2 \pm1.2)\times 10^{-9}$ \\
\hline
$\alpha=\{0, 2/3, 2\}$       &  $(-7.9 \pm11.4)\times 10^{-8}$ &   $(9.5 \pm11.3)\times 10^{-8}$ &   $(-1.8 \pm1.5)\times 10^{-8}$ &                                - &                                   - \\
$\alpha=\{0, 2/3, 3\}$       &   $(-3.2 \pm8.3)\times 10^{-8}$ &    $(3.5 \pm7.0)\times 10^{-8}$ &                               - &    $(-2.9 \pm2.7)\times 10^{-9}$ &                                   - \\
$\alpha=\{0, 2/3, 4\}$       &  $(-9.9 \pm69.8)\times 10^{-9}$ &    $(1.3 \pm5.6)\times 10^{-8}$ &                               - &                                - &      $(-6.2 \pm6.1)\times 10^{-10}$ \\
$\alpha=\{0, 2, 3\}$         &  $(-1.0 \pm30.0)\times 10^{-9}$ &                               - &   $(8.3 \pm22.6)\times 10^{-9}$ &    $(-4.1 \pm6.4)\times 10^{-9}$ &                                   - \\
$\alpha=\{0, 2, 4\}$         &   $(4.3 \pm25.0)\times 10^{-9}$ &                               - &   $(1.0 \pm12.9)\times 10^{-9}$ &                                - &     $(-5.9 \pm10.5)\times 10^{-10}$ \\
$\alpha=\{0, 3, 4\}$         &   $(6.5 \pm17.8)\times 10^{-9}$ &                               - &                               - &  $(-1.8 \pm81.2)\times 10^{-10}$ &     $(-4.6 \pm23.2)\times 10^{-10}$ \\
$\alpha=\{2/3, 2, 3\}$       &                               - &   $(1.7 \pm38.8)\times 10^{-9}$ &   $(6.1 \pm34.4)\times 10^{-9}$ &    $(-3.6 \pm8.4)\times 10^{-9}$ &                                   - \\
$\alpha=\{2/3, 2, 4\}$       &                               - &   $(7.7 \pm30.1)\times 10^{-9}$ &  $(-1.7 \pm19.5)\times 10^{-9}$ &                                - &     $(-4.5 \pm12.7)\times 10^{-10}$ \\
$\alpha=\{2/3, 3, 4\}$       &                               - &   $(8.1 \pm18.3)\times 10^{-9}$ &                               - &   $(-1.9 \pm10.4)\times 10^{-9}$ &  $(-5.6 \pm280.6)\times 10^{-11}$ \\
$\alpha=\{2, 3, 4\}$         &                               - &                               - &    $(1.6 \pm2.8)\times 10^{-8}$ &    $(-1.1 \pm2.5)\times 10^{-8}$ &        $(1.5 \pm4.9)\times 10^{-9}$ \\
\hline
$\alpha=\{0, 2/3, 2, 3\}$    &   $(-3.2 \pm3.5)\times 10^{-7}$ &    $(4.1 \pm4.6)\times 10^{-7}$ &   $(-1.2 \pm1.5)\times 10^{-7}$ &     $(1.9 \pm2.6)\times 10^{-8}$ &                                   - \\
$\alpha=\{0, 2/3, 2, 4\}$    &   $(-2.5 \pm2.7)\times 10^{-7}$ &    $(3.0 \pm3.3)\times 10^{-7}$ &   $(-6.8 \pm7.6)\times 10^{-8}$ &                                - &        $(2.1 \pm3.1)\times 10^{-9}$ \\
$\alpha=\{0, 2/3, 3, 4\}$    &   $(-1.5 \pm1.8)\times 10^{-7}$ &    $(1.6 \pm1.9)\times 10^{-7}$ &                               - &    $(-2.3 \pm2.8)\times 10^{-8}$ &        $(4.5 \pm6.3)\times 10^{-9}$ \\
$\alpha=\{0, 2, 3, 4\}$      &   $(-4.0 \pm6.3)\times 10^{-8}$ &                               - &    $(7.6 \pm9.9)\times 10^{-8}$ &    $(-4.7 \pm6.2)\times 10^{-8}$ &       $(7.1 \pm10.2)\times 10^{-9}$ \\
$\alpha=\{2/3, 2, 3, 4\}$    &                               - &   $(-5.5 \pm9.6)\times 10^{-8}$ &   $(9.9 \pm14.7)\times 10^{-8}$ &    $(-5.4 \pm7.8)\times 10^{-8}$ &       $(7.8 \pm12.0)\times 10^{-9}$ \\
\hline
$\alpha=\{0, 2/3, 2, 3, 4\}$ &   $(-6.9 \pm8.9)\times 10^{-7}$ &   $(9.8 \pm13.6)\times 10^{-7}$ &   $(-4.3 \pm7.1)\times 10^{-7}$ &     $(1.3 \pm2.6)\times 10^{-7}$ &       $(-1.4 \pm3.0)\times 10^{-8}$ \\
\bottomrule
\end{tabular}
}
\caption{Estimators from the multi-components analysis in the 20-100 Hz band for the different combinations of the five spectral indices. Horizontal lines divide the table in regions where a fixed number of components is considered for the analysis.}
\label{tab:PL_analysis_estimators_table}
\end{table*}

\begin{table*}
\scriptsize
\makebox[\textwidth][c]{
\begin{tabular}{llllll|lll}
\toprule
{} &  $\Omega_{0}^{95\%}$ & $\Omega_{2/3}^{95\%}$ &  $\Omega_{2}^{95\%}$ &  $\Omega_{3}^{95\%}$ &  $\Omega_{4}^{95\%}$ & $K_{\rm CBC}^{95\%}$ & $K_{\rm magnetars}^{95\%}$ & $K_{\rm r-modes}^{95\%}$ \\
\midrule
$\alpha=\{0\}$               &  $1.6\times 10^{-8}$ &                     - &                    - &                    - &                    - &                    - &                          - &                        - \\
$\alpha=\{2/3\}$             &                    - &   $1.2\times 10^{-8}$ &                    - &                    - &                    - &   $5.1\times 10^{4}$ &                          - &                        - \\
$\alpha=\{2\}$               &                    - &                     - &  $4.1\times 10^{-9}$ &                    - &                    - &                    - &                          - &       $1.3\times 10^{0}$ \\
$\alpha=\{3\}$               &                    - &                     - &                    - &  $1.5\times 10^{-9}$ &                    - &                    - &                          - &                        - \\
$\alpha=\{4\}$               &                    - &                     - &                    - &                    - &  $4.5\times10^{-10}$ &                    - &       $1.4\times 10^{-12}$ &                        - \\
\hline
$\alpha=\{0, 2/3\}$          &  $1.3\times 10^{-8}$ &   $9.5\times 10^{-9}$ &                    - &                    - &                    - &   $4.2\times 10^{4}$ &                          - &                        - \\
$\alpha=\{0, 2\}$            &  $1.4\times 10^{-8}$ &                     - &  $3.5\times 10^{-9}$ &                    - &                    - &                    - &                          - &       $1.3\times 10^{0}$ \\
$\alpha=\{0, 3\}$            &  $1.5\times 10^{-8}$ &                     - &                    - &  $1.3\times 10^{-9}$ &                    - &                    - &                          - &                        - \\
$\alpha=\{0, 4\}$            &  $1.5\times 10^{-8}$ &                     - &                    - &                    - &  $4.1\times10^{-10}$ &                    - &       $1.3\times 10^{-12}$ &                        - \\
$\alpha=\{2/3, 2\}$          &                    - &   $1.0\times 10^{-8}$ &  $3.5\times 10^{-9}$ &                    - &                    - &   $4.9\times 10^{4}$ &                          - &       $1.3\times 10^{0}$ \\
$\alpha=\{2/3, 3\}$          &                    - &   $1.0\times 10^{-8}$ &                    - &  $1.3\times 10^{-9}$ &                    - &   $4.6\times 10^{4}$ &                          - &                        - \\
$\alpha=\{2/3, 4\}$          &                    - &   $1.0\times 10^{-8}$ &                    - &                    - &  $4.1\times10^{-10}$ &   $4.9\times 10^{4}$ &       $1.3\times 10^{-12}$ &                        - \\
$\alpha=\{2, 3\}$            &                    - &                     - &  $3.7\times 10^{-9}$ &  $1.3\times 10^{-9}$ &                    - &                    - &                          - &       $1.3\times 10^{0}$ \\
$\alpha=\{2, 4\}$            &                    - &                     - &  $3.9\times 10^{-9}$ &                    - &  $4.0\times10^{-10}$ &                    - &       $1.3\times 10^{-12}$ &       $1.3\times 10^{0}$ \\
$\alpha=\{3, 4\}$            &                    - &                     - &                    - &  $1.3\times 10^{-9}$ &  $4.0\times10^{-10}$ &                    - &       $1.3\times 10^{-12}$ &                        - \\
\hline
$\alpha=\{0, 2/3, 2\}$       &  $1.2\times 10^{-8}$ &   $8.4\times 10^{-9}$ &  $3.1\times 10^{-9}$ &                    - &                    - &   $4.3\times 10^{4}$ &                          - &       $1.3\times 10^{0}$ \\
$\alpha=\{0, 2/3, 3\}$       &  $1.2\times 10^{-8}$ &   $8.8\times 10^{-9}$ &                    - &  $1.2\times 10^{-9}$ &                    - &   $3.9\times 10^{4}$ &                          - &                        - \\
$\alpha=\{0, 2/3, 4\}$       &  $1.3\times 10^{-8}$ &   $8.9\times 10^{-9}$ &                    - &                    - &  $3.9\times10^{-10}$ &   $4.2\times 10^{4}$ &       $1.2\times 10^{-12}$ &                        - \\
$\alpha=\{0, 2, 3\}$         &  $1.3\times 10^{-8}$ &                     - &  $3.2\times 10^{-9}$ &  $1.1\times 10^{-9}$ &                    - &                    - &                          - &       $1.3\times 10^{0}$ \\
$\alpha=\{0, 2, 4\}$         &  $1.4\times 10^{-8}$ &                     - &  $3.3\times 10^{-9}$ &                    - &  $3.8\times10^{-10}$ &                    - &       $1.3\times 10^{-12}$ &       $1.3\times 10^{0}$ \\
$\alpha=\{0, 3, 4\}$         &  $1.4\times 10^{-8}$ &                     - &                    - &  $1.2\times 10^{-9}$ &  $3.8\times10^{-10}$ &                    - &       $1.2\times 10^{-12}$ &                        - \\
$\alpha=\{2/3, 2, 3\}$       &                    - &   $9.0\times 10^{-9}$ &  $3.2\times 10^{-9}$ &  $1.1\times 10^{-9}$ &                    - &   $4.6\times 10^{4}$ &                          - &       $1.3\times 10^{0}$ \\
$\alpha=\{2/3, 2, 4\}$       &                    - &   $9.4\times 10^{-9}$ &  $3.3\times 10^{-9}$ &                    - &  $3.8\times10^{-10}$ &   $4.9\times 10^{4}$ &       $1.3\times 10^{-12}$ &       $1.3\times 10^{0}$ \\
$\alpha=\{2/3, 3, 4\}$       &                    - &   $9.5\times 10^{-9}$ &                    - &  $1.2\times 10^{-9}$ &  $3.8\times10^{-10}$ &   $4.4\times 10^{4}$ &       $1.2\times 10^{-12}$ &                        - \\
$\alpha=\{2, 3, 4\}$         &                    - &                     - &  $3.4\times 10^{-9}$ &  $1.2\times 10^{-9}$ &  $3.7\times10^{-10}$ &                    - &       $1.3\times 10^{-12}$ &       $1.3\times 10^{0}$ \\
\hline
$\alpha=\{0, 2/3, 2, 3\}$    &  $1.2\times 10^{-8}$ &   $7.8\times 10^{-9}$ &  $2.9\times 10^{-9}$ &  $1.1\times 10^{-9}$ &                    - &   $3.9\times 10^{4}$ &                          - &       $1.3\times 10^{0}$ \\
$\alpha=\{0, 2/3, 2, 4\}$    &  $1.2\times 10^{-8}$ &   $8.2\times 10^{-9}$ &  $2.8\times 10^{-9}$ &                    - &  $3.7\times10^{-10}$ &   $4.1\times 10^{4}$ &       $1.2\times 10^{-12}$ &       $1.3\times 10^{0}$ \\
$\alpha=\{0, 2/3, 3, 4\}$    &  $1.2\times 10^{-8}$ &   $8.3\times 10^{-9}$ &                    - &  $1.1\times 10^{-9}$ &  $3.6\times10^{-10}$ &   $3.9\times 10^{4}$ &       $1.2\times 10^{-12}$ &                        - \\
$\alpha=\{0, 2, 3, 4\}$      &  $1.3\times 10^{-8}$ &                     - &  $2.9\times 10^{-9}$ &  $1.1\times 10^{-9}$ &  $3.6\times10^{-10}$ &                    - &       $1.2\times 10^{-12}$ &       $1.3\times 10^{0}$ \\
$\alpha=\{2/3, 2, 3, 4\}$    &                    - &   $8.7\times 10^{-9}$ &  $3.0\times 10^{-9}$ &  $1.1\times 10^{-9}$ &  $3.5\times10^{-10}$ &   $4.4\times 10^{4}$ &       $1.2\times 10^{-12}$ &       $1.3\times 10^{0}$ \\
\hline
$\alpha=\{0, 2/3, 2, 3, 4\}$ &  $1.1\times 10^{-8}$ &   $7.6\times 10^{-9}$ &  $2.8\times 10^{-9}$ &  $1.0\times 10^{-9}$ &  $3.4\times10^{-10}$ &   $3.8\times 10^{4}$ &       $1.2\times 10^{-12}$ &       $1.3\times 10^{0}$ \\
\bottomrule
\end{tabular}
}
\caption{95\% Bayesian upper limits from the multi-components analysis in the 20-100 Hz band on $\Omega_{\alpha}^{95\%}$ (to the left of the vertical line) and the related astrophysical parameters $K_{i}^{95\%}$ (to the right of the vertical line) for the different combinations of the five spectral indices. Horizontal lines divide the table in regions where a fixed number of components is considered for the analysis. The constraints on $\Omega_{\alpha}^{95\%}$ have been obtained using uniform prior between 0 and $10^{-6.5}$. The constraints on $K_{i}^{95\%}$ have been obtained using log-uniform prior, see main text. $K_{\rm CBC}$ units are $M_{\odot}\, {\rm Gpc^{-3}\, yr^{-1}}$, $K_{\rm r-modes}$ is dimensionless, and $K_{\rm magnetars}$ is expressed in $\mathrm{T}^{-1}$.}
\label{tab:unified_UL_table}
\end{table*}

\subsection{Astrophysical parameters}

\begin{figure*}
    \centering
    \includegraphics[width=0.48\textwidth]{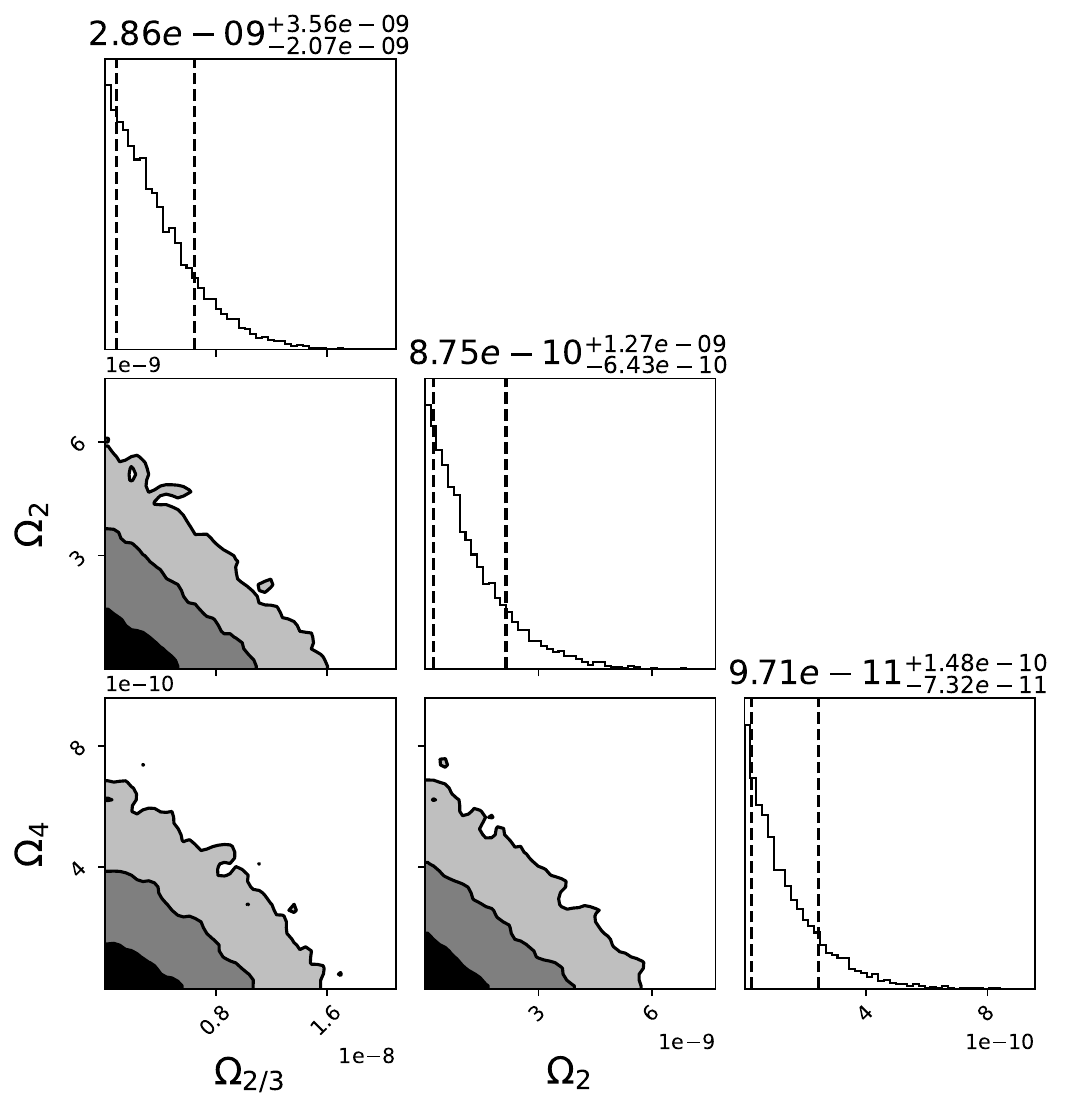}
    \includegraphics[width=0.49\textwidth]{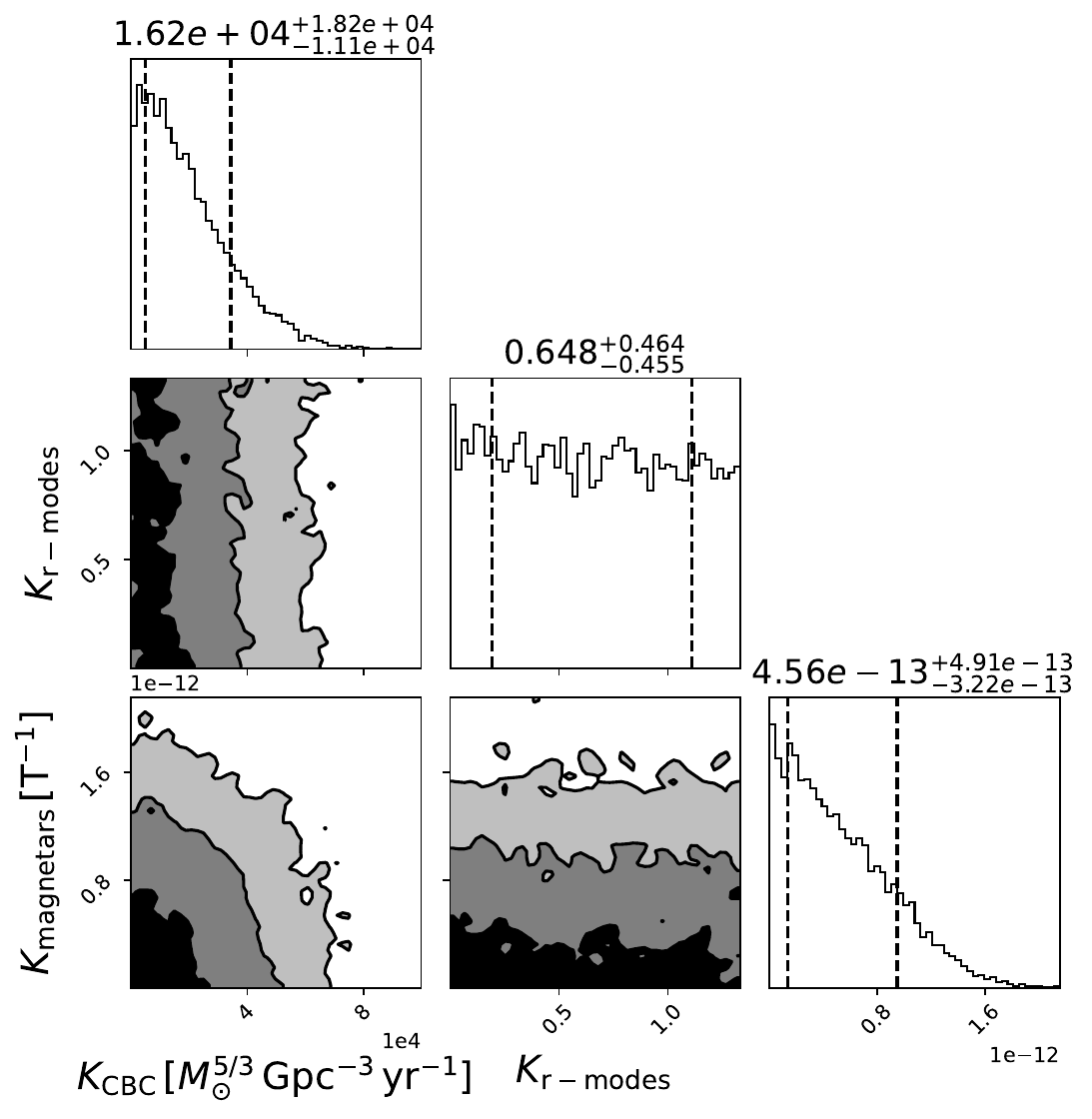}
    \caption{Results of the parameter estimation for the $\alpha = 2/3,\, 2,\,4$ combination in the $20-100$ Hz band for power-law $\Omega_{\alpha}$ (left panel) and the corresponding astrophysical parameters (right panel). Contour plots show the $1\sigma$, $2\sigma$, and $3\sigma$ credible areas (black, grey, light grey, respectively). The dashed black lines in the histogram panels delimit the $1\sigma$ region of the estimated parameters.}
    \label{fig:Astro_analysis_PE_plot}
\end{figure*}

The previous results for the $\Omega_{\alpha}$ can be reinterpreted and translated into constraints on the ensemble properties of the SGWBs of interest, namely from CBCs ($\alpha=2/3$), r-mode instabilities ($\alpha = 2$), and magnetars ($\alpha =4$). This can be done in different ways. Suppose the SGWB consists of just one component, and just one ensemble property $\expval{\theta^{c}}$ characterises the background. In that case, it is possible to build an estimator $\widehat{\expval{\theta^{c}}}$ for such quantity (see \cite{ellipticity_paper_DeLillo:2022blw, glitch_paper_DeLillo:2022dau} as examples of this approach) and the associated likelihood to use to draw the Bayesian upper limits. Alternatively, suppose multiple components are present, each having more than one ensemble property. In that case, building the estimator for each property (fixed a component) is generally impossible. The solution is to change the models for $\Omega_{\alpha}$ in the likelihood \eqref{eq:likelihood_Omega_joint} with the ones in equation \eqref{eq:Om_a_ensemble_properties_PE} when performing the parameter estimation or constraining the ensemble properties.

In this work, we have opted for the second approach. Given the absence of signal, we have been able only to constrain the ensemble properties for each component, namely $K_{\CBC}, \, K_{\rm r-modes}, \, K_{\rm magnetars}$.
The corresponding $95\%$-confidence Bayesian upper limits are reported in table \ref{tab:unified_UL_table}. To obtain such constraints, we have used uniform priors, with $K_{\CBC}\in [0,\,  10^{7}]\, M_{\odot}^{5/3}\, {\rm Gpc^{-3}\, yr^{-1}}$, $K_{\rm r-modes} \in [10^{-13},\, 4/3]$, and $ K_{\rm magnetars} \in [0, \, 10^{-10}] \, {\rm T^{-1}}$, motivated by the limits derived in \cite{GWTC-3_pop_KAGRA:2021duu}, \cite{r-modes_Sa_Tome:2006hn}, and \cite{magnetars_2013PhRvD..87d2002W}, respectively. By inspection, we observe that the limits on $K_{\rm r-modes}$ and $K_{\rm magnetars}$ mildly depend on the number of components considered. In contrast, $K_{\CBC}$ limits oscillate more, with variations between 5\%-25\% with respect to the value inferred from the single-component analysis. The reference values that we quote as results of this work are the ones from the $\alpha = \{2/3, \, 2,\, 4\}$ combination, namely $K_{\CBC}\leq 4.9 \times10^{4} M_{\odot}^{5/3}\, {\rm Gpc^{-3}\, yr^{-1}}$, $K_{\rm r-modes}\leq 1.3$, and $ K_{\rm magnetars} \leq 1.3 \times 10^{-12}\, \mathrm{T}^{-1}$. The choice of not including $\alpha =0, \, 3$ for these reference values comes from the injection study presented in the appendix \ref{sec:injection_study}, assuming only these three components are present/dominant. %Are these constraints strong enough to have relevant implications on the astrophysical populations? Let's examine each of them individually.
We examine each of these constraints individually to check whether they are strong enough to have relevant implications for astrophysical populations.

The CBC parameter $K_{\rm CBC}$ is actually a bound over the sum of the products $r_{0,\, j}\langle\mathcal{M}_{c}^{5/3}\rangle_{j}$ ($j= \BBH,\,\BNS,\, \NSBH$) involving the local rates and the average $5/3$ power of the chirp masses of the individual $\BBH, \, \BNS ,\, {\rm and}\, \NSBH$ populations. Its interpretation and comparison with the recent limits for CBC population from the GWTC-3 catalogue \cite{GWTC-3_pop_KAGRA:2021duu} are not straightforward. This is a limit of the multi-component analysis under the assumption of power-law energy density spectra for $\Omega_{\gw}(f)$, which does not allow to remove the degeneracy between SGWBs having the same spectral indices but different ensemble parameters. Further studies about how to break such degeneracy will be the subject of future works, aiming to bring the multi-component analysis beyond the simple power-law assumption.

The r-mode-instability parameter $K_{\rm r-modes} = \langle(K+2)^{-1}\rangle$
can be approximately converted in a limit over $\expval{K}$ by taking its inverse and subtracting 2. Doing this for the present upper limit leads to a lower limit for $\expval{K} \gtrsim -1.23$. This value is right above $-5/4$, the minimal value that $K$ can assume according to \cite{r-modes_Sa_Tome:2006hn} and corresponds to the maximum value of $\Omega_{\gw,\, {\rm r-modes}}$. This limit reflects that not all r-mode instabilities happen with this extreme value for $K$, usually assumed when doing estimates of the intensity of this kind of SGWB, but does not exclude that individual events in that configuration from happening. Still, the weakness of the constraint does not allow to draw major implications for these phenomena, where uncertainties still dominate many parameters.

The magnetar parameter $K_{\rm magnetars} = \sqrt{\expval{\varepsilon^{2}}\,\expval{B^{-2}}}$ suffers from the degeneracy between the average square ellipticity and the average (inverse) square of the (poloidal) magnetic field of the magnetar population. If the magnetar population had an average magnetic field around $10^{10}\, (10^{11})$ T, the corresponding limits on (square root of the) average (squared) ellipticity would be $\varepsilon \leq 1.3\times 10^{-2} \, (10^{-1})$. Following equations (13) and (14) from \cite{magnetars_2013PhRvD..87d2002W}, this, in turn, would imply the limits on the distortion parameter $\beta\leq 3.5 \times 10^{5}\, (10^{4})$ for a poloidal-dominated field configuration and the dimensionless parameter $k\leq1.3\times 10^{6}\, (10^{5})$ for a twisted-toroidal field configuration (assuming $\lambda_{\rm magnetars} = 9\times 10^{-4}\, M_{\odot}$). However, these constraints are not informative and, together with the other astrophysical uncertainties in the magnetar population, do not allow to draw further implications for the geometry of individual magnetars or on the equation of state of the magnetar population. 

\section{\label{sec:conclusions} Conclusions}

In this work, we have performed a search for a Gaussian, stationary, unpolarised, isotropic stochastic gravitational-wave background with a power-law spectrum in energy density, relaxing the assumption that only one component can be present at a time. We have performed such a search assuming different SGWBs to be present at a time, with spectral indices $\alpha =\{ 0, \, 2/3, \, 2,\, 3, \, 4\}$ and for all possible combinations, in a $20-100$ Hz frequency band. Then, we have inferred the implications for astrophysical SGWBs from compact binary coalescences, young-pulsar r-mode instabilities, and magnetars, corresponding to the spectral indices $\alpha =\{2/3, \, 2,\, 4\}$.

The analysis has not shown evidence of any SGWB, so we have set upper limits on $\Omega_{\alpha}$ at 25-Hz reference frequency for every $\alpha$ combination. We recover the limits from \cite{o3_iso_KAGRA:2021kbb} in the single-index case when performing the analysis in the $20-1726$ Hz range. When multiple components are present, the limits become more stringent. The derivation of the implications for the ensemble properties of the astrophysical SGWB from CBCs, r-mode instabilities, and magnetars results in constraints over $K_{\CBC}$, $K_{\rm r-modes}$, and $K_{\rm magnetars}$, respectively. The bounds are not informative in the case of r-modes, and are not competitive with the existing ones in the case of CBCs and magnetars. This fact is further reflected by their mild oscillations in value when considering different $\alpha$ combinations.

The results obtained in this paper may not include additional information compared to the existing ones in the literature, given the weakness of the SGWB components in the data. However, as shown and discussed in the injection study in appendix \ref{sec:injection_study}, the method employed here will be fundamental to avoid bias and overestimation of the components when getting closer to a detection. The same injection study also highlights the necessity of generalising this method to the case where the SGWB cannot be described by a simple power law and, consequently, introduces bias in the search presented in this work if the signal is strong enough to generate a power excess in the data or be detected. This will allow us to avoid limits on the frequency range used in the search where the power-law regime is no longer applicable. In addition to that, adapting this method to any frequency dependence will also allow to easily employ it for directional and targeted searches, even in the case where match filtering for the SGWB of interest is used (see as example \cite{targeted_search_ellipticity_Agarwal:2022lvk}). We reserve such generalisations for future works.

\section{\label{Acknowledgements} Acknowledgements}

This material is based upon work supported by NSF's LIGO Laboratory which is a major facility fully funded by the National Science Foundation. This research has used data obtained from the Gravitational Wave Open Science Center (\url{https://www.gw-openscience.org/data/})~\citep{gwosc:LIGOScientific:2019lzm, gwosc:KAGRA:2023pio}, a service of LIGO Laboratory, the LIGO Scientific Collaboration, the Virgo Collaboration, and KAGRA. 
The authors are grateful for computational resources provided by the
LIGO Laboratory and supported by NSF Grants PHY-0757058 and PHY-0823459, and they acknowledge the use of the supercomputing facilities of the Universit\'e catholique de Louvain (CISM/UCLouvain) and the Consortium des \'Equipements de Calcul Intensif en F\'ed\'eration Wallonie Bruxelles (C\'ECI), funded by the Fond de la Recherche Scientifique de Belgique (F.R.S.-FNRS) under convention 2.5020.11 and by the Walloon Region. The authors gratefully acknowledge the support of the NSF, STFC, INFN and CNRS for the provision of computational resources. This article has a LIGO document number LIGO-P2300335.

The authors are grateful to Thomas Callister for carefully reading the manuscript and providing valuable comments. They also thank Tania Regimbau for useful discussions and suggestions about astrophysical models in the early stages of this work.

F.D.L. is supported by a FRIA (Fonds pour la formation à la Recherche dans l'Industrie et dans l'Agriculture) Grant of the Belgian Fund for Research, F.R.S.-FNRS (Fonds de la Recherche Scientifique-FNRS). J.S is supported by a Actions de Recherche Concertées (ARC) grant.

We have used {\tt{numpy}} \cite{numpy}, {\tt{scipy}} \cite{scipy}, and {\tt{matplotlib}} \cite{matplotlib} packages to handle data and produce figures of this work. We have performed Bayesian inference by using the {\tt numpyro} package \cite{numpyro}.

\bibliography{joint_analysis_astro}% Produces the bibliography via BibTeX.

\appendix
\section{Injection studies}
\label{sec:injection_study}
To further validate the multi-component search method, we have performed several injection studies in the frequency domain, going beyond the scenarios present in \cite{Joint_iso_methods_Parida:2015fma}, where this method was first introduced. The validation process has allowed us to better understand how to interpret the results of the multi-component analysis and compare them with the single-component ones, testing the limit of this formalism for determining the ensemble properties of an SGWB with a frequency-power-law energy density spectrum. We present here the injection studies that summarise the main cases of interest.

\subsection{Power-law injections: five equal-intensity SGWBs}
\label{sec:equal_PL_injections}
The first set of injections we present and discuss reproduce a scenario where several components are present, are sufficiently intense to be detectable with O3 sensitivity, and have all the same intensity. The injected signals are characterised by $\{\Omega_{\alpha}=1 \times 10^{-6},\, \alpha = 0, 2/3, 2, 3, 4\}$ at 25 Hz, and the analysis is performed in the $20-1726$ Hz band for all possible $\alpha$ combinations. 
The choice of this broad frequency range allows for a better sensitivity of the search for higher $\alpha$ (the O3 public data sets \cite{O3_iso_data} are also available in this range).

\subsubsection{Signal-only case using O3 sensitivity}
In this toy model, whose results are summarised in table \ref{tab:Equal_1e-6_PL_injections_estimator_table_no_noise}, we do not add any random noise to the signal, though we still account for the O3-noise power spectral densities (PSDs). The absence of additional noise allows us to isolate the bias that the single-index analysis introduces in the estimates (and in the error bars) of the individual components.

We observe that all the results of the component estimation in table \ref{tab:Equal_1e-6_PL_injections_estimator_table_no_noise} show an excess associated with a signal, independent of the combination of the spectral indices. However, only the joint estimation that considers all five indices recovers the injections correctly. This is further confirmed by the parameter estimation (PE) results that are shown in figure \ref{fig:strong_PL_injections_Oms_PE_plot} (up to negligible numerical errors in the estimators due to the inversion of the coupling matrix $\Gamma'_{\alpha \alpha'}$). In all the other cases, the estimation leads to heavily biased results, with the highest bias in the case of single-component analysis. This can be understood as the power of the ignored components spreading and getting absorbed by the components considered in the analysis.

\begin{table*}
\scriptsize
%\makebox[\textwidth][c]{
\resizebox{\textwidth}{!}{
\begin{tabular}{llllll}
\toprule
{} &   $\hat{\Omega}_{0}=1\times 10^{-6}$ & $\hat{\Omega}_{2/3}=1\times 10^{-6}$ &  $\hat{\Omega}_{2}=1\times 10^{-6}$ &  $\hat{\Omega}_{3}=1\times 10^{-6}$ &  $\hat{\Omega}_{4}=1\times 10^{-6}$ \\
\midrule
$\alpha=\{0\}$               &    $(1.9420 \pm0.0008)\times 10^{-5}$ &                                    - &                                   - &                                   - &                                   - \\
$\alpha=\{2/3\}$             &                                    - &   $(1.8066 \pm0.0006)\times 10^{-5}$ &                                   - &                                   - &                                   - \\
$\alpha=\{2\}$               &                                    - &                                    - &  $(1.2519 \pm0.0002)\times 10^{-5}$ &                                   - &                                   - \\
$\alpha=\{3\}$               &                                    - &                                    - &                                   - &   $(5.6650 \pm0.0008)\times 10^{-6}$ &                                   - \\
$\alpha=\{4\}$               &                                    - &                                    - &                                   - &                                   - &  $(1.2262 \pm0.0002)\times 10^{-6}$ \\
\hline
$\alpha=\{0, 2/3\}$          &  $(-1.5621 \pm0.0004)\times 10^{-4}$ &   $(1.3276 \pm0.0003)\times 10^{-4}$ &                                   - &                                   - &                                   - \\
$\alpha=\{0, 2\}$            &    $(-3.595 \pm0.001)\times 10^{-5}$ &                                    - &  $(2.1765 \pm0.0004)\times 10^{-5}$ &                                   - &                                   - \\
$\alpha=\{0, 3\}$            &    $(-9.516 \pm0.009)\times 10^{-6}$ &                                    - &                                   - &   $(6.205 \pm0.001)\times 10^{-6}$ &                                   - \\
$\alpha=\{0, 4\}$            &     $(7.665 \pm0.008)\times 10^{-6}$ &                                    - &                                   - &                                   - &  $(1.1886 \pm0.0002)\times 10^{-6}$ \\
$\alpha=\{2/3, 2\}$          &                                    - &    $(-3.874 \pm0.001)\times 10^{-5}$ &   $(2.7240 \pm0.0005)\times 10^{-5}$ &                                   - &                                   - \\
$\alpha=\{2/3, 3\}$          &                                    - &    $(-9.227 \pm0.007)\times 10^{-6}$ &                                   - &    $(6.513 \pm0.001)\times 10^{-6}$ &                                   - \\
$\alpha=\{2/3, 4\}$          &                                    - &     $(6.305 \pm0.006)\times 10^{-6}$ &                                   - &                                   - &  $(1.1696 \pm0.0002)\times 10^{-6}$ \\
$\alpha=\{2, 3\}$            &                                    - &                                    - &   $(-9.976 \pm0.005)\times 10^{-6}$ &    $(8.751 \pm0.002)\times 10^{-6}$ &                                   - \\
$\alpha=\{2, 4\}$            &                                    - &                                    - &    $(3.521 \pm0.003)\times 10^{-6}$ &                                   - &  $(1.0827 \pm0.0002)\times 10^{-6}$ \\
$\alpha=\{3, 4\}$            &                                    - &                                    - &                                   - &    $(1.819 \pm0.002)\times 10^{-6}$ &    $(9.141 \pm0.003)\times 10^{-7}$ \\
\hline
$\alpha=\{0, 2/3, 2\}$       &    $(2.253 \pm0.001)\times 10^{-4}$ &  $(-2.5345 \pm0.0009)\times 10^{-4}$ &    $(5.086 \pm0.001)\times 10^{-5}$ &                                   - &                                   - \\
$\alpha=\{0, 2/3, 3\}$       &     $(9.418 \pm0.007)\times 10^{-5}$ &    $(-8.525 \pm0.005)\times 10^{-5}$ &                                   - &    $(8.154 \pm0.002)\times 10^{-6}$ &                                   - \\
$\alpha=\{0, 2/3, 4\}$       &    $(-2.188 \pm0.005)\times 10^{-5}$ &     $(2.278 \pm0.004)\times 10^{-5}$ &                                   - &                                   - &  $(1.1288 \pm0.0002)\times 10^{-6}$ \\
$\alpha=\{0, 2, 3\}$         &     $(2.918 \pm0.002)\times 10^{-5}$ &                                    - &   $(-2.584 \pm0.001)\times 10^{-5}$ &  $(1.2001 \pm0.0003)\times 10^{-5}$ &                                   - \\
$\alpha=\{0, 2, 4\}$         &        $(-9.7 \pm0.2)\times 10^{-7}$ &                                    - &    $(3.839 \pm0.006)\times 10^{-6}$ &                                   - &  $(1.0745 \pm0.0002)\times 10^{-6}$ \\
$\alpha=\{0, 3, 4\}$         &      $(3.35 \pm0.01)\times 10^{-6}$ &                                    - &                                   - &    $(1.385 \pm0.002)\times 10^{-6}$ &    $(9.722 \pm0.004)\times 10^{-7}$ \\
$\alpha=\{2/3, 2, 3\}$       &                                    - &     $(3.603 \pm0.002)\times 10^{-5}$ &   $(-3.527 \pm0.002)\times 10^{-5}$ &  $(1.3262 \pm0.0003)\times 10^{-5}$ &                                   - \\
$\alpha=\{2/3, 2, 4\}$       &                                    - &       $(-1.20 \pm0.02)\times 10^{-6}$ &    $(4.076 \pm0.008)\times 10^{-6}$ &                                   - &  $(1.0708 \pm0.0003)\times 10^{-6}$ \\
$\alpha=\{2/3, 3, 4\}$       &                                    - &     $(2.904 \pm0.009)\times 10^{-6}$ &                                   - &    $(1.255 \pm0.002)\times 10^{-6}$ &    $(9.848 \pm0.004)\times 10^{-7}$ \\
$\alpha=\{2, 3, 4\}$         &                                    - &                                    - &    $(2.599 \pm0.008)\times 10^{-6}$ &      $(5.32 \pm0.04)\times 10^{-7}$ &  $(1.0289 \pm0.0005)\times 10^{-6}$ \\
\hline
$\alpha=\{0, 2/3, 2, 3\}$    &    $(-1.525 \pm0.001)\times 10^{-4}$ &     $(2.045 \pm0.002)\times 10^{-4}$ &   $(-7.061 \pm0.004)\times 10^{-5}$ &  $(1.7369 \pm0.0005)\times 10^{-5}$ &                                   - \\
$\alpha=\{0, 2/3, 2, 4\}$    &       $(1.18 \pm0.01)\times 10^{-5}$ &       $(-1.30 \pm0.01)\times 10^{-5}$ &      $(5.68 \pm0.02)\times 10^{-6}$ &                                   - &   $(1.054 \pm0.0003)\times 10^{-6}$ \\
$\alpha=\{0, 2/3, 3, 4\}$    &       $(-1.50 \pm0.08)\times 10^{-6}$ &       $(4.18 \pm0.07)\times 10^{-6}$ &                                   - &    $(1.202 \pm0.004)\times 10^{-6}$ &    $(9.898 \pm0.005)\times 10^{-7}$ \\
$\alpha=\{0, 2, 3, 4\}$      &       $(1.82 \pm0.03)\times 10^{-6}$ &                                    - &      $(1.29 \pm0.02)\times 10^{-6}$ &      $(9.44 \pm0.07)\times 10^{-7}$ &  $(1.0027 \pm0.0006)\times 10^{-6}$ \\
$\alpha=\{2/3, 2, 3, 4\}$    &                                    - &       $(2.21 \pm0.03)\times 10^{-6}$ &        $(6.6 \pm0.3)\times 10^{-7}$ &    $(1.063 \pm0.009)\times 10^{-6}$ &     $(9.970 \pm0.007)\times 10^{-7}$ \\
\hline
$\alpha=\{0, 2/3, 2, 3, 4\}$ &         $(1.0 \pm0.2)\times 10^{-6}$ &        $(1.0 \pm0.3)\times 10^{-6}$ &       $(1.00 \pm0.07)\times 10^{-6}$ &       $(1.00 \pm0.02)\times 10^{-6}$ &     $(1.0000 \pm0.0009)\times 10^{-6}$ \\
\bottomrule
\end{tabular}
}
\caption{Estimators from the multi-components analysis in the $20-1726$ Hz band for the different combinations of the five spectral indices for the signal-only, power-law injection data set, with $\Omega_{\alpha}=10^{-6}$. Horizontal lines divide the table in regions where a fixed number of components is considered for the analysis.}
\label{tab:Equal_1e-6_PL_injections_estimator_table_no_noise}
\end{table*}

This bias can be easily quantified. Let us consider the case where one is performing the joint estimation assuming only the spectral indices $\alpha$ while missing the spectral indices $\alpha_{m}$. Then the SGWB can be written as ($A\equiv \{\alpha,\, \alpha_{m}\}$)
\begin{equation}
    \label{eq:extra_components}
    \Omega_{\gw}(f) = \sum_{A}\Omega_{A} w_{A}(f) = \sum_{\alpha}\Omega_{\alpha}w_{\alpha}(f) + \sum_{\alpha_{m}}\Omega_{\alpha_{m}} w_{\alpha_{m}}(f) \, .
\end{equation}
One should then perform the analysis by including both $\alpha$ and $\alpha_{m}$, building the dirty map $X_{A}=(X_{\alpha}, \, X_{\alpha_{m}})$ and the Fisher matrix $\Gamma_{AA'}$ ($A=\{\alpha, \alpha_{m}\}$), and then deriving the unbiased estimator $\hat{\Omega}_{A} = (\Gamma^{-1})_{AA'}X_{A'}$, such that $\expval{\hat{\Omega}_{A}} = \Omega_{A}$. If one restricts the search to $\alpha$ only, one has (summation over repeated indices is implicit)
\begin{align}
    \hat{\Omega}_{\alpha}^{(\rm unbiased)} &= (\Gamma^{-1})_{\alpha\alpha'}X_{\alpha'} - (\Gamma^{-1})_{\alpha \alpha'}\Gamma_{\alpha'\alpha'_{m}}\,\Omega_{\alpha'_{m}} \nonumber\\
    &= \hat{\Omega}_{\alpha}^{(\rm biased)} - (\Gamma^{-1})_{\alpha \alpha'}\Gamma_{\alpha'\alpha'_{m}}\,\Omega_{\alpha'_{m}}, 
\end{align}
where $\hat{\Omega}_{\alpha}^{(\rm biased)}$ is the (\textit{biased}) estimator from equation \eqref{eq:Om_a_estimator_preconditioned} that one gets when ``missing'' the components $\Omega_{\alpha_{m}}$. This means that the bias is
\begin{equation}
    \label{eq:bias_ignoring_components}
    b_{\alpha} = (\Gamma^{-1})_{\alpha \alpha'}\Gamma_{\alpha'\alpha'_{m}}\, \Omega_{\alpha'_{m}}.
\end{equation}
This bias can be positive or negative, and it is not known a priori, given one has not considered the $\Omega_{\alpha_{m}}$ in this scenario. A naive scaling of its absolute value is given by the number of components in the analysis times the (unknown) number of missed components\footnote{This is confirmed from the results in table \ref{tab:Equal_1e-6_PL_injections_estimator_table_no_noise}. The most biased results appear in the analyses for two and three components that miss three or two components, respectively.}. %A situation where multiple components are ignored appears to be highly undesirable and should be avoided.
A situation where this product is maximum (and, naively, the bias is maximum) appears to be highly undesirable and should be avoided. 
Nonetheless (as it is further discussed in section \ref{sec:dominant_PL_injection}), the ultimate discriminant for the importance of this bias is the relative magnitude of the missed components $\Omega_{\alpha_{m}}$ with respect to the ones of the $\Omega_{\alpha}$ components that one considers in the analysis. The above expression for the bias reduces to the one reported in \cite{Joint_iso_methods_Parida:2015fma} in the case of the estimator of the single component analysis.

\subsubsection{Signal on top of O3 data}
\begin{figure*}
    \centering
    \includegraphics[width= 0.49\textwidth]{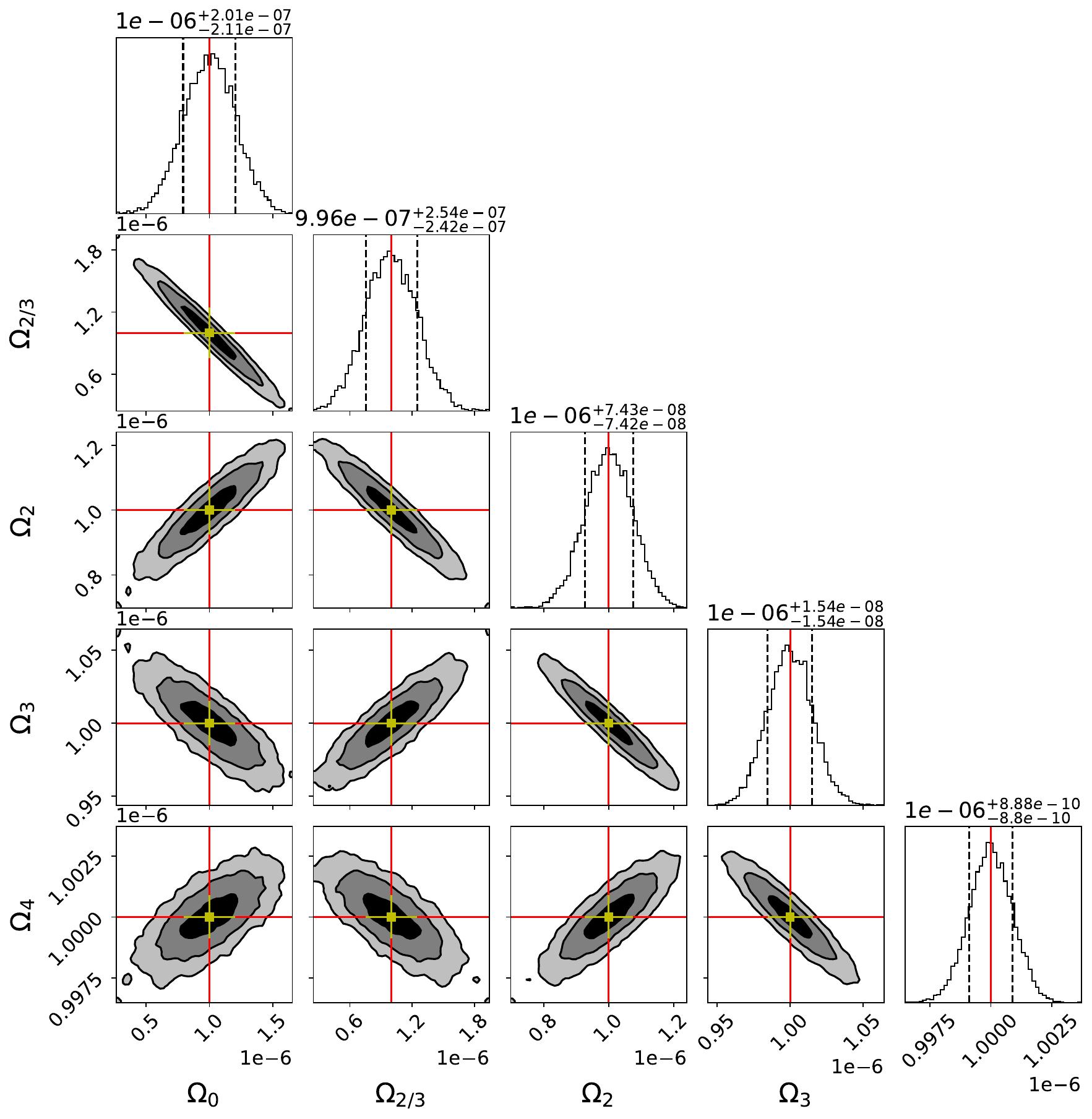}
    \includegraphics[width= 0.49\textwidth]{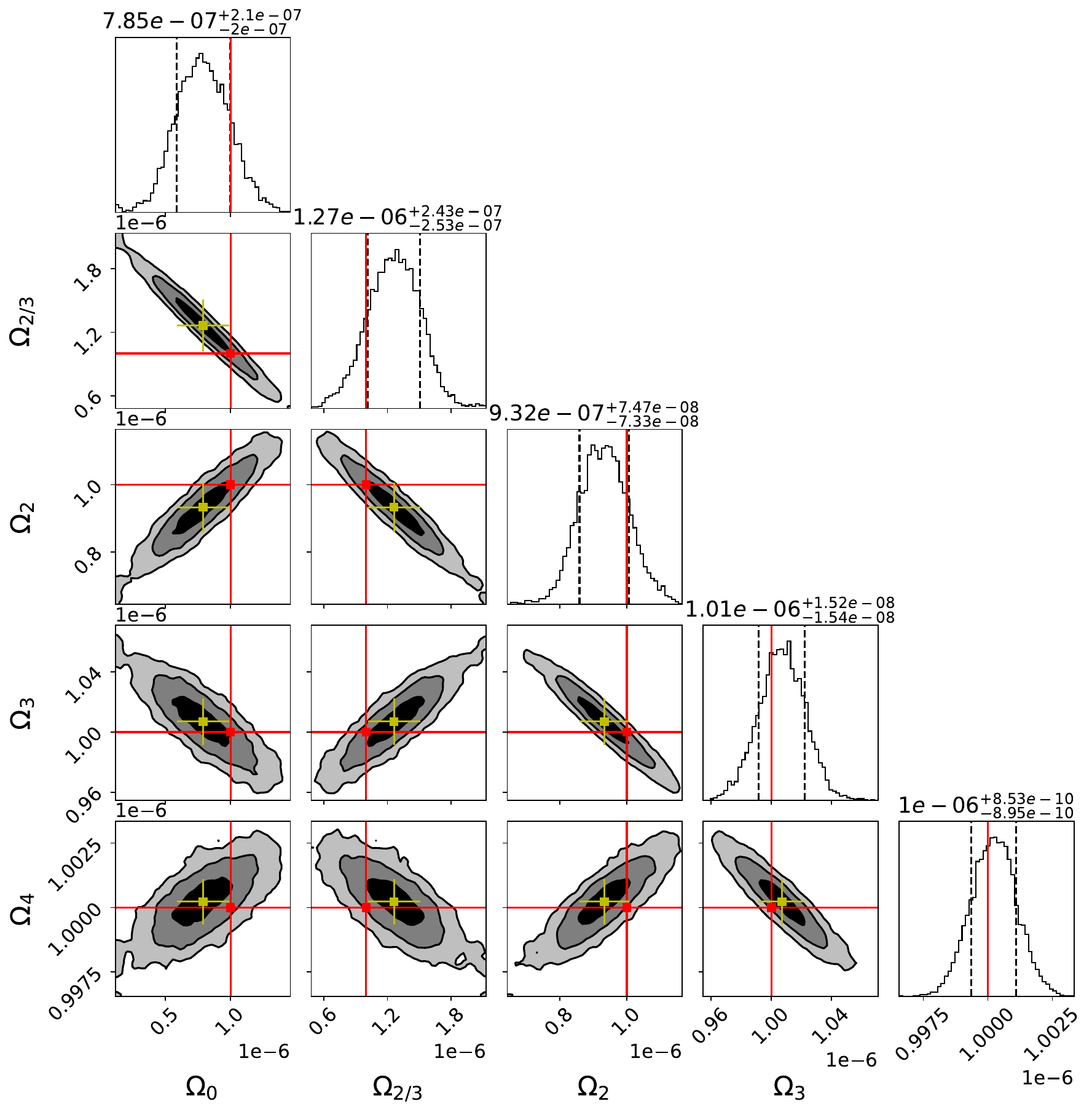}
    \caption{Parameter estimation results for the set of the power-law injections in the $20-1726$ Hz frequency range, with $\Omega_{\alpha} = 1\times 10^{-6}$, $\alpha= {0,\, 2/3, \, 2,\, 3,\, 4}$, for the signal-only case (left panel) and on top of O3 data (right panel). Contour plots show the $1\sigma$, $2\sigma$, and $3\sigma$ credible areas (black, grey, light grey, respectively). The red lines denote the injected values, while the yellow error bars represent the $1\sigma$ uncertainty of the $\Omega_{\alpha}$ estimators from the joint analysis. The dashed black lines in the histogram panels delimit the $1\sigma$ region of the estimated parameters.}
    \label{fig:strong_PL_injections_Oms_PE_plot}
\end{figure*}

To get closer to a more realistic scenario, we inject the same signal on top of O3 data, which in turn acts as a source of unknown noise.
The results for all the possible $\alpha$ combinations in terms of estimators $\Omega_{\alpha}$ are summarised in table \ref{tab:Strong_PL_injections_estimator_table}, while the PE results, together with the injected values and the resulting estimators, for the five-index multi-component analysis of the five indices are illustrated in figure \ref{fig:strong_PL_injections_Oms_PE_plot}.

Analogously to the case where only the signal is present, the injections are recovered only when all five indices are considered in the analysis. However, in contrast to the previous toy model case, we observe that the recovery (within 1-sigma) and the estimators appear to be biased even when all indices enter into the analysis. This is related to the fact that we are actually not considering one component in the analysis, namely the spectrum from O3 data $\hat{\Omega}_{\rm O3}(f)$. As a consequence, $\hat{\Omega}_{\rm O3}(f)$ is acting as a source of an unknown noise that does not follow a power-law and gets absorbed in the components we consider from time to time in the analysis.
%This is a consequence that here there is actually one component that we are not considering in the analysis, namely the spectrum from O3 data $\hat{\Omega}_{\rm O3}(f)$, which is acting as a source of an unknown noise, not following a power law, that gets absorbed in the components we consider from time to time. 
A source of unknown noise mimicking the signal is a challenge 
%that must be faced 
even in the usual single-component analysis; in that case, the bias will be larger than (or at best comparable with) the one observed in the multi-component analysis that includes all the components but the noise. In addition to that, the presence of correlation between two different spectral indices (visible in the contour plots in figure \ref{fig:strong_PL_injections_Oms_PE_plot}) may lead to degeneracy or additional bias in the recovery when the signals are not sufficiently intense and/or distant enough in the spectral-index space.
%We observe that, for any combination, estimators and PE results in tables \ref{tab:Strong_PL_injections_estimator_table} and \ref{tab:Strong_PL_injections_PE_table} make clear that there are excesses associated with a signal. However, only the combination considering all the five indices does recover the injection correctly, within 1-sigma uncertainty, while the other combinations lead to heavily biased results. In addition to this, by looking at figure \ref{fig:strong_PL_injections_Oms_PE_plot}, it is possible to notice the presence of correlation between two different spectral indices in the contour plots, which may lead to degeneracy or bias in the recovery when the signals are not sufficiently intense and/or distant enough in the spectral-index space. Similar results hold for detectable injections where the intensities of the injected $\Omega_{\alpha}$ differ for every $\alpha$, even by orders of magnitude. 

\begin{table*}
\scriptsize
%\makebox[\textwidth][c]{
\resizebox{\textwidth}{!}{
\begin{tabular}{llllll}
\toprule
{} &   $\hat{\Omega}_{0}=1\times 10^{-6}$ & $\hat{\Omega}_{2/3}=1\times 10^{-6}$ &  $\hat{\Omega}_{2}=1\times 10^{-6}$ &  $\hat{\Omega}_{3}=1\times 10^{-6}$ &  $\hat{\Omega}_{4}=1\times 10^{-6}$ \\
\midrule
$\alpha=\{0\}$               &   $(1.9421 \pm0.0008)\times 10^{-5}$ &                                    - &                                   - &                                   - &                                   - \\
$\alpha=\{2/3\}$             &                                    - &   $(1.8066 \pm0.0006)\times 10^{-5}$ &                                   - &                                   - &                                   - \\
$\alpha=\{2\}$               &                                    - &                                    - &  $(1.2517 \pm0.0002)\times 10^{-5}$ &                                   - &                                   - \\
$\alpha=\{3\}$               &                                    - &                                    - &                                   - &  $(5.6644 \pm0.0008)\times 10^{-6}$ &                                   - \\
$\alpha=\{4\}$               &                                    - &                                    - &                                   - &                                   - &  $(1.2263 \pm0.0002)\times 10^{-6}$ \\
\hline
$\alpha=\{0, 2/3\}$          &  $(-1.5616 \pm0.0004)\times 10^{-4}$ &   $(1.3272 \pm0.0003)\times 10^{-4}$ &                                   - &                                   - &                                   - \\
$\alpha=\{0, 2\}$            &    $(-3.593 \pm0.001)\times 10^{-5}$ &                                    - &  $(2.1759 \pm0.0004)\times 10^{-5}$ &                                   - &                                   - \\
$\alpha=\{0, 3\}$            &     $(-9.510 \pm0.009)\times 10^{-6}$ &                                    - &                                   - &    $(6.204 \pm0.001)\times 10^{-6}$ &                                   - \\
$\alpha=\{0, 4\}$            &     $(7.666 \pm0.008)\times 10^{-6}$ &                                    - &                                   - &                                   - &  $(1.1887 \pm0.0002)\times 10^{-6}$ \\
$\alpha=\{2/3, 2\}$          &                                    - &    $(-3.873 \pm0.001)\times 10^{-5}$ &  $(2.7233 \pm0.0005)\times 10^{-5}$ &                                   - &                                   - \\
$\alpha=\{2/3, 3\}$          &                                    - &    $(-9.223 \pm0.007)\times 10^{-6}$ &                                   - &    $(6.512 \pm0.001)\times 10^{-6}$ &                                   - \\
$\alpha=\{2/3, 4\}$          &                                    - &     $(6.305 \pm0.006)\times 10^{-6}$ &                                   - &                                   - &  $(1.1696 \pm0.0002)\times 10^{-6}$ \\
$\alpha=\{2, 3\}$            &                                    - &                                    - &   $(-9.975 \pm0.005)\times 10^{-6}$ &     $(8.750 \pm0.002)\times 10^{-6}$ &                                   - \\
$\alpha=\{2, 4\}$            &                                    - &                                    - &    $(3.518 \pm0.003)\times 10^{-6}$ &                                   - &  $(1.0828 \pm0.0002)\times 10^{-6}$ \\
$\alpha=\{3, 4\}$            &                                    - &                                    - &                                   - &    $(1.817 \pm0.002)\times 10^{-6}$ &    $(9.146 \pm0.003)\times 10^{-7}$ \\
\hline
$\alpha=\{0, 2/3, 2\}$       &    $(2.2532 \pm0.001)\times 10^{-4}$ &  $(-2.5342 \pm0.0009)\times 10^{-4}$ &    $(5.085 \pm0.001)\times 10^{-5}$ &                                   - &                                   - \\
$\alpha=\{0, 2/3, 3\}$       &     $(9.423 \pm0.007)\times 10^{-5}$ &    $(-8.528 \pm0.005)\times 10^{-5}$ &                                   - &    $(8.154 \pm0.002)\times 10^{-6}$ &                                   - \\
$\alpha=\{0, 2/3, 4\}$       &    $(-2.181 \pm0.005)\times 10^{-5}$ &     $(2.273 \pm0.004)\times 10^{-5}$ &                                   - &                                   - &   $(1.1290 \pm0.0002)\times 10^{-6}$ \\
$\alpha=\{0, 2, 3\}$         &     $(2.921 \pm0.002)\times 10^{-5}$ &                                    - &   $(-2.585 \pm0.001)\times 10^{-5}$ &  $(1.2004 \pm0.0003)\times 10^{-5}$ &                                   - \\
$\alpha=\{0, 2, 4\}$         &        $(-9.5 \pm0.2)\times 10^{-7}$ &                                    - &    $(3.828 \pm0.006)\times 10^{-6}$ &                                   - &  $(1.0748 \pm0.0002)\times 10^{-6}$ \\
$\alpha=\{0, 3, 4\}$         &      $(3.36 \pm0.01)\times 10^{-6}$ &                                    - &                                   - &    $(1.381 \pm0.002)\times 10^{-6}$ &    $(9.729 \pm0.004)\times 10^{-7}$ \\
$\alpha=\{2/3, 2, 3\}$       &                                    - &     $(3.607 \pm0.002)\times 10^{-5}$ &   $(-3.529 \pm0.002)\times 10^{-5}$ &  $(1.3266 \pm0.0003)\times 10^{-5}$ &                                   - \\
$\alpha=\{2/3, 2, 4\}$       &                                    - &      $(-1.17 \pm0.02)\times 10^{-6}$ &     $(4.060 \pm0.008)\times 10^{-6}$ &                                   - &  $(1.0713 \pm0.0003)\times 10^{-6}$ \\
$\alpha=\{2/3, 3, 4\}$       &                                    - &     $(2.917 \pm0.009)\times 10^{-6}$ &                                   - &     $(1.250 \pm0.002)\times 10^{-6}$ &    $(9.856 \pm0.004)\times 10^{-7}$ \\
$\alpha=\{2, 3, 4\}$         &                                    - &                                    - &    $(2.611 \pm0.008)\times 10^{-6}$ &      $(5.24 \pm0.04)\times 10^{-7}$ &  $(1.0299 \pm0.0005)\times 10^{-6}$ \\
\hline
$\alpha=\{0, 2/3, 2, 3\}$    &    $(-1.528 \pm0.001)\times 10^{-4}$ &     $(2.048 \pm0.002)\times 10^{-4}$ &   $(-7.069 \pm0.004)\times 10^{-5}$ &   $(1.7380 \pm0.0005)\times 10^{-5}$ &                                   - \\
$\alpha=\{0, 2/3, 2, 4\}$    &       $(1.17 \pm0.01)\times 10^{-5}$ &      $(-1.29 \pm0.01)\times 10^{-5}$ &      $(5.64 \pm0.02)\times 10^{-6}$ &                                   - &  $(1.0547 \pm0.0003)\times 10^{-6}$ \\
$\alpha=\{0, 2/3, 3, 4\}$    &      $(-1.55 \pm0.08)\times 10^{-6}$ &       $(4.23 \pm0.07)\times 10^{-6}$ &                                   - &    $(1.196 \pm0.004)\times 10^{-6}$ &    $(9.907 \pm0.005)\times 10^{-7}$ \\
$\alpha=\{0, 2, 3, 4\}$      &       $(1.82 \pm0.03)\times 10^{-6}$ &                                    - &       $(1.3 \pm0.02)\times 10^{-6}$ &      $(9.37 \pm0.07)\times 10^{-7}$ &  $(1.0036 \pm0.0006)\times 10^{-6}$ \\
$\alpha=\{2/3, 2, 3, 4\}$    &                                    - &       $(2.21 \pm0.03)\times 10^{-6}$ &        $(6.7 \pm0.3)\times 10^{-7}$ &    $(1.057 \pm0.009)\times 10^{-6}$ &    $(9.979 \pm0.007)\times 10^{-7}$ \\
\hline
$\alpha=\{0, 2/3, 2, 3, 4\}$ &         $(7.9 \pm2.0)\times 10^{-7}$ &         $(1.3 \pm0.2)\times 10^{-6}$ &        $(9.3 \pm0.7)\times 10^{-7}$ &      $(1.01 \pm0.02)\times 10^{-6}$ &  $(1.0002 \pm0.0009)\times 10^{-6}$ \\
\bottomrule
\end{tabular}
}
\caption{Estimators from the multi-components analysis in the $20-1726$ Hz band for the different combinations of the five spectral indices for power-law injection data set on top of the O3 data, with $\Omega_{\alpha}=10^{-6}$. Horizontal lines divide the table in regions where a fixed number of components is considered for the analysis.}
\label{tab:Strong_PL_injections_estimator_table}
\end{table*}

\subsection{Power-law injections: three different-intensity SGWBs}
\label{sec:dominant_PL_injection}
The previous set of injections assumes that all the components are equally contributing to a given frequency. However, it may be more realistic to assume that the amplitude of the components differs even by orders of magnitude. %This second set of injections tests the multi-component method exactly in such a case.
In this second injection set, we reproduce the possible scenario at 25 Hz in figure \ref{fig:landscape_plot} for the CBC, r-mode, and magnetar SGWBs, and we inject the signals corresponding to the spectral indices $\alpha = 2/3, \, 2,\, 4$, such that $\Omega_{2} = 10^{-3}\times \Omega_{2/3}$ and $\Omega_{4} = 10^{-5}\times \Omega_{2/3}$ at 25 Hz. We still perform the analysis for all the combinations of spectral indices $\{\alpha = 0, 2/3, 2, 3, 4\}$ to investigate the impact of the different choices that are discussed below.

\subsubsection{Signal-only case with O3 rescaled sensitivity}
In this toy model, we again do not add any random noise to the data, but we inject only the signals, with $\Omega_{2/3} = 1 \times 10^{-9}$, $\Omega_{2} = 1 \times 10^{-12}$, and $\Omega_{4} = 1 \times 10^{-14}$. In addition to that, we also scale down the O3 PSDs in such a way that the SGWBs are detectable. The results of the joint estimation are collected in table \ref{tab:2/3_dominant_PL_injections_estimator_table_no_noise} for every spectral indices combination, and figure \ref{fig:dominant_PL_injection_Om_PE_plots} shows the PE results for the combination $\alpha = 2/3, \, 2, \, 4$.
Again, the estimator central values perfectly match the injections only if all the right indices corresponding to the injections are included in the analysis. However, there are some noticeable differences from the injection datasets with the equal-amplitude components that are presented in section \ref{sec:equal_PL_injections}.

\begin{table*}
\scriptsize
%\makebox[\textwidth][c]{
\resizebox{\textwidth}{!}{
\begin{tabular}{llllll}
\toprule
{} &                                    $\hat{\Omega}_{0}=0$ & $\hat{\Omega}_{2/3}=1\times 10^{-9}$ &  $\hat{\Omega}_{2}=1\times 10^{-12}$ &                                   $\hat{\Omega}_{3}=0$ &  $\hat{\Omega}_{4}=1\times 10^{-14}$ \\
\midrule
$\alpha=\{0\}$               &                      $(1.32561 \pm 0.00008)\times 10^{-9}$ &                                    - &                                    - &                                                      - &                                    - \\
$\alpha=\{2/3\}$             &                                                       - &   $(1.00219 \pm 0.00006)\times 10^{-9}$ &                                    - &                                                      - &                                    - \\
$\alpha=\{2\}$               &                                                       - &                                    - &  $(3.8106 \pm0.0002)\times 10^{-10}$ &                                                      - &                                    - \\
$\alpha=\{3\}$               &                                                       - &                                    - &                                    - &                    $(9.2225 \pm0.0008)\times 10^{-11}$ &                                    - \\
$\alpha=\{4\}$               &                                                       - &                                    - &                                    - &                                                      - &    $(9.041 \pm0.002)\times 10^{-12}$ \\
\hline
$\alpha=\{0, 2/3\}$          &                          $(-8.7 \pm0.4)\times 10^{-12}$ &   $(1.0086 \pm0.0003)\times 10^{-9}$ &                                    - &                                                      - &                                    - \\
$\alpha=\{0, 2\}$            &                      $(1.0306 \pm0.0001)\times 10^{-9}$ &                                    - &  $(1.1598 \pm0.0004)\times 10^{-10}$ &                                                      - &                                    - \\
$\alpha=\{0, 3\}$            &                      $(1.21771 \pm0.00009)\times 10^{-9}$ &                                    - &                                    - &                     $(2.314 \pm0.001)\times 10^{-11}$ &                                    - \\
$\alpha=\{0, 4\}$            &                      $(1.29924 \pm0.00008)\times 10^{-9}$ &                                    - &                                    - &                                                      - &    $(2.666 \pm0.002)\times 10^{-12}$ \\
$\alpha=\{2/3, 2\}$          &                                                       - &    $(9.996 \pm0.001)\times 10^{-10}$ &      $(1.22 \pm0.05)\times 10^{-12}$ &                                                      - &                                    - \\
$\alpha=\{2/3, 3\}$          &                                                       - &   $(1.00116 \pm0.00007)\times 10^{-9}$ &                                    - &                          $(2.4 \pm0.1)\times 10^{-13}$ &                                    - \\
$\alpha=\{2/3, 4\}$          &                                                       - &   $(1.00184 \pm0.00006)\times 10^{-9}$ &                                    - &                                                      - &        $(3.4 \pm0.2)\times 10^{-14}$ \\
$\alpha=\{2, 3\}$            &                                                       - &                                    - &  $(7.0273 \pm0.0005)\times 10^{-10}$ &                   $(-1.2513 \pm0.0002)\times 10^{-10}$ &                                    - \\
$\alpha=\{2, 4\}$            &                                                       - &                                    - &  $(4.6276 \pm0.0003)\times 10^{-10}$ &                                                      - &   $(-9.831 \pm0.002)\times 10^{-12}$ \\
$\alpha=\{3, 4\}$            &                                                       - &                                    - &                                    - &                     $(1.948 \pm0.0002)\times 10^{-10}$ &  $(-2.438 \pm0.0003)\times 10^{-11}$ \\
\hline
$\alpha=\{0, 2/3, 2\}$       &                           $(2.0 \pm1.0)\times 10^{-12}$ &    $(9.977 \pm0.009)\times 10^{-10}$ &        $(1.4 \pm0.1)\times 10^{-12}$ &                                                      - &                                    - \\
$\alpha=\{0, 2/3, 3\}$       &                          $(-2.5 \pm0.7)\times 10^{-12}$ &   $(1.0032 \pm0.0005)\times 10^{-9}$ &                                    - &                          $(2.0 \pm0.2)\times 10^{-13}$ &                                    - \\
$\alpha=\{0, 2/3, 4\}$       &                          $(-5.9 \pm0.5)\times 10^{-12}$ &   $(1.0063 \pm0.0004)\times 10^{-9}$ &                                    - &                                                      - &        $(2.3 \pm0.2)\times 10^{-14}$ \\
$\alpha=\{0, 2, 3\}$         &                       $(8.887 \pm0.002)\times 10^{-10}$ &                                    - &    $(2.196 \pm0.001)\times 10^{-10}$ &                     $(-2.613 \pm0.003)\times 10^{-11}$ &                                    - \\
$\alpha=\{0, 2, 4\}$         &                         $(9.800 \pm0.002)\times 10^{-10}$ &                                    - &  $(1.4192 \pm0.0006)\times 10^{-10}$ &                                                      - &   $(-1.555 \pm0.002)\times 10^{-12}$ \\
$\alpha=\{0, 3, 4\}$         &                     $(1.1619 \pm0.0001)\times 10^{-9}$ &                                    - &                                    - &                      $(4.404 \pm0.002)\times 10^{-11}$ &   $(-4.217 \pm0.004)\times 10^{-12}$ \\
$\alpha=\{2/3, 2, 3\}$       &                                                       - &   $(1.0003 \pm0.0002)\times 10^{-9}$ &        $(6.4 \pm1.7)\times 10^{-13}$ &                          $(1.2 \pm0.3)\times 10^{-13}$ &                                    - \\
{$\alpha=\{2/3, 2, 4\}$}       &                                                       - &      {$(1.0000 \pm0.0002)\times 10^{-9}$} &       {$(1.00 \pm0.08)\times 10^{-12}$} &                                                      - &        {$(1.0 \pm0.3)\times 10^{-14}$} \\
$\alpha=\{2/3, 3, 4\}$       &                                                       - &   $(1.00106 \pm0.00009)\times 10^{-9}$ &                                    - &                          $(2.9 \pm0.2)\times 10^{-13}$ &       $(-8.5 \pm3.8)\times 10^{-15}$ \\
$\alpha=\{2, 3, 4\}$         &                                                       - &                                    - &  $(8.7956 \pm0.0008)\times 10^{-10}$ &                    $(-2.40705_ \pm0.0004)\times 10^{-10}$ &  $(1.4469 \pm0.0005)\times 10^{-11}$ \\
\hline
$\alpha=\{0, 2/3, 2, 3\}$    &                          $(-1.5 \pm1.5)\times 10^{-12}$ &     $(1.002 \pm0.002)\times 10^{-9}$ &        $(2.8 \pm3.8)\times 10^{-13}$ &                          $(1.6 \pm0.5)\times 10^{-13}$ &                                    - \\
$\alpha=\{0, 2/3, 2, 4\}$    &  $(-0.00 \pm1.14)\times 10^{-11}$ &       $(1.000 \pm0.001)\times 10^{-9}$ &       $(1.00 \pm0.17)\times 10^{-12}$ &                                                      - &        $(1.0 \pm0.3)\times 10^{-14}$ \\
$\alpha=\{0, 2/3, 3, 4\}$    &                          $(-2.5 \pm0.8)\times 10^{-12}$ &   $(1.0032 \pm0.0007)\times 10^{-9}$ &                                    - &                          $(2.0 \pm0.4)\times 10^{-13}$ &      $(-2.3 \pm45.9)\times 10^{-16}$ \\
$\alpha=\{0, 2, 3, 4\}$      &                       $(8.159 \pm0.003)\times 10^{-10}$ &                                    - &    $(2.919 \pm0.002)\times 10^{-10}$ &                     $(-5.559 \pm0.007)\times 10^{-11}$ &    $(2.671 \pm0.006)\times 10^{-12}$ \\
$\alpha=\{2/3, 2, 3, 4\}$    &                                                       - &      $(1.0000 \pm0.0003)\times 10^{-9}$ &       $(1.00 \pm0.29)\times 10^{-12}$ &                          $(0.0 \pm8.7)\times 10^{-14}$ &       $(1.00 \pm0.66)\times 10^{-14}$ \\
\hline
$\alpha=\{0, 2/3, 2, 3, 4\}$ &  $(-0.00 \pm2.02)\times 10^{-12}$ &       $(1.000 \pm0.002)\times 10^{-9}$ &        $(1.0 \pm0.7)\times 10^{-12}$ &  $(0.00 \pm1.54)\times 10^{-13}$ &       $(1.00 \pm0.89)\times 10^{-14}$ \\
\bottomrule
\end{tabular}
}
\caption{Estimators from the multi-components analysis in the $20-1726$ Hz band for different spectral-index combinations for the dominant-component, signal-only, power-law injection data set, with $\Omega_{2/3} = 1 \times 10^{-9}$, $\Omega_{2} = 1 \times 10^{-12}$, and $\Omega_{4} = 1 \times 10^{-14}$. Horizontal lines divide the table in regions where a fixed number of components is considered for the analysis.}
\label{tab:2/3_dominant_PL_injections_estimator_table_no_noise}
\end{table*}

First, we observe that the biases in the estimators for the dominant component are, at most, in the order of a few percentages compared to the injected value among all cases. This may suggest that even single-component analysis may be effective when there is a component that is dominant over all the others by multiple orders of magnitude\footnote{However, this may no longer be true if $\Omega_{2} = 10^{-2}-10^{-1} \times \Omega_{2/3}$, since the bias in the single-component estimator increases by a factor of $10-100$, passing from $0.22\%$ in table \ref{tab:2/3_dominant_PL_injections_estimator_table_no_noise} to $2.2\%-22\%$.}. However, we stress that only the joint analysis allows for the correct recovery of the injected value, and the bias may grow when injecting (and ignoring) other components in spite of their weakness compared to the $\Omega_{2/3}$ intensity. In addition to that, the other subdominant components are heavily biased by the dominant one, as long as they are not considered all together. This is another reason to use multi-component analysis even in the presence of a dominant component when that cannot be completely subtracted from the data.

Second, we can examine how different $\alpha$ combinations affect the estimators of the $\Omega_{0}$ and $\Omega_{3}$ components, which are not present in the data. The multi-component search is capable of recognising the zero amplitudes of these components when they are considered in the analysis together with the injected $\Omega_{2/3}$, $\Omega_{2}$, and $\Omega_{4}$ ones. This means that in contrast to when we are missing some components, performing the analysis with extra components does not introduce any bias in the estimators.
%as long as all the components containing the signal are included in the analysis.

%Third and last, we can notice the impact of the non-injected components in the case where they are considered in the analysis, but not all the injected ones are. This is just equivalent to the case where we are missing some signal components, and hence, the joint estimation of the components is biased, including the zero-amplitude components acquiring non-zero values. 

Third and last, we can notice the impact of the non-injected components when we are missing some signal components. Similarly to the signal-only case in section \ref{sec:equal_PL_injections}, the joint estimation of the components is biased, with the additional complication of the zero-amplitude components acquiring non-zero values.  

\subsubsection{On top of O3 data}
\begin{figure*}
    \centering
    \includegraphics[width = 0.49\textwidth]{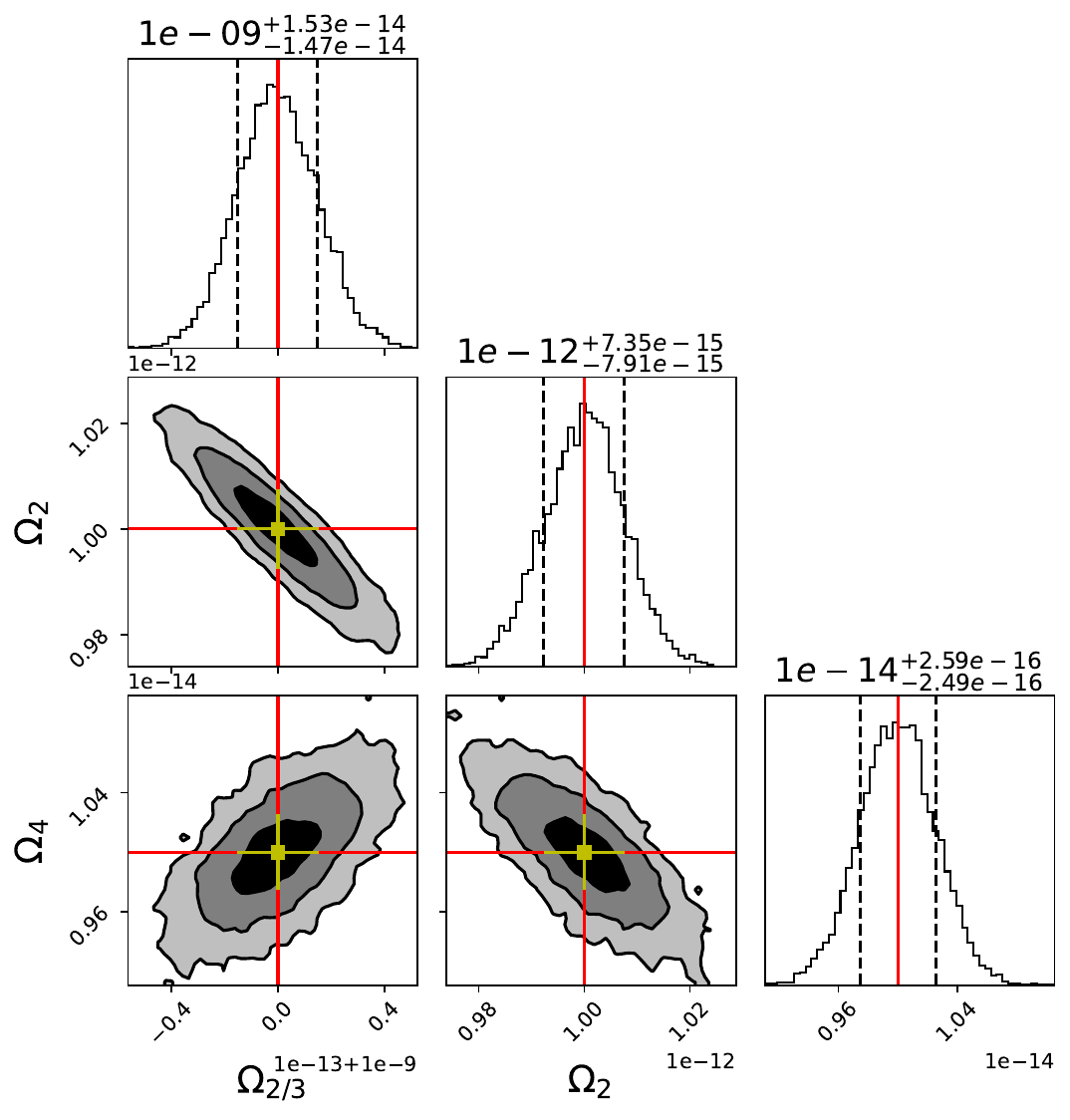}
    \includegraphics[width = 0.49\textwidth]{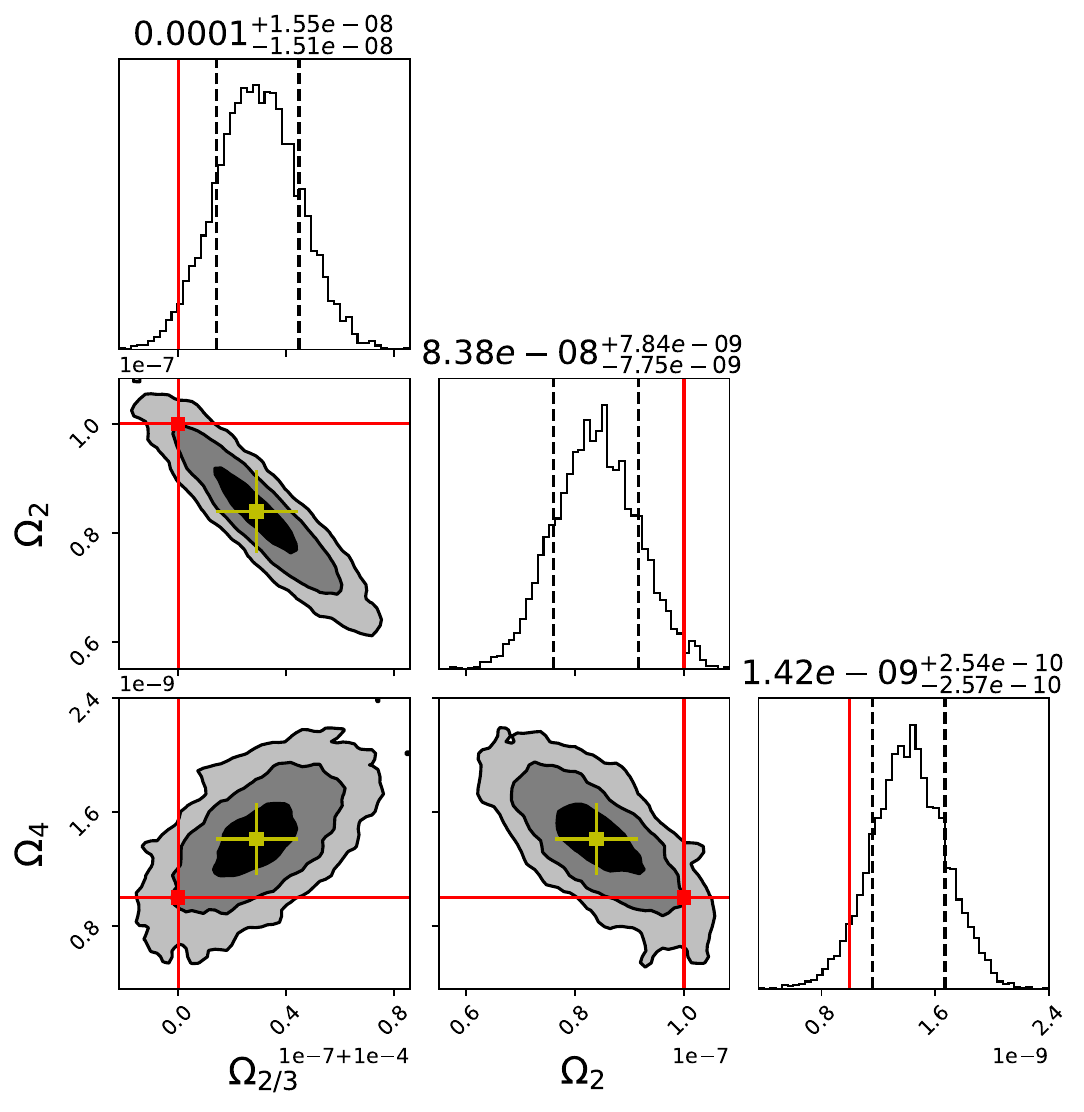}
    \caption{Parameter estimation results for $\alpha= {2/3, \, 2,\, 4}$ from the two sets of the power-law injections in the $20-1726$ Hz frequency range, with $\Omega_{2} = 10^{-3}\times \Omega_{2/3}$, $\Omega_{4} = 10^{-5}\times \Omega_{2/3}$, for the signal-only ($\Omega_{2/3}=10^{-9}$, left) and the on-top-of-O3-data ($\Omega_{2/3}=10^{-4}$, right) injections. Contour plots show the $1\sigma$, $2\sigma$, and $3\sigma$ credible areas (black, grey, light grey, respectively). The red lines denote the injected values, while the yellow error bars represent the $1\sigma$ uncertainty of the $\Omega_{\alpha}$ estimators from the joint analysis. The dashed black lines in the histogram panels delimit the $1\sigma$ region of the estimated parameters.}
    \label{fig:dominant_PL_injection_Om_PE_plots}
\end{figure*}

In this injection study, we inject the signals on top of O3 data, using the extremely high values $\Omega_{2/3}=1 \times 10^{-4}$, $\Omega_{2} = 1 \times 10^{-7}$, and $\Omega_{4} = 1 \times 10^{-9}$ to guarantee the detectability of the signal. We limit ourselves to illustrating the PE plots in the joint $\alpha = 2/3, \, 2, \, 4$ case in figure \ref{fig:dominant_PL_injection_Om_PE_plots}. The recovery (within 2-sigma) of the injections is biased due to the presence of the O3 noise $\Omega_{\rm O3}$ that is not accounted for in the analysis. However, in spite of being visually evident, the bias in the dominant component is tiny, being roughly $0.01\%$ of the injected value.

\subsection{Astrophysical injections}
The third and last set of injections mimics a scenario where the injected $\Omega_{\gw, \, i}(f)$ ($i = 1, 2,\dotsc, \mathrm{N_{components}}$) are no longer a power-law in the full search frequency band $20-1726$ Hz, as in the case of the astrophysical SGWBs described in section \ref{sec:astro_SGWB}. In addition to that, the overall intensity of the $\Omega_{\gw, \,i}(f)$ is related to the choice of the ensemble properties of the SGWBs that we aim to constrain using the results from the multi-component analysis. In this injection data set, we have injected $\Omega_{\gw, \, i}(f)$, with $i = $ BNS, r-modes, magnetars, and population properties $K_{\rm BNS} \simeq 7.91 \times 10^{5}\, {\rm M_{\odot}^{5/3}\, Gpc^{-3}\, yr^{-1}}$, $K_{\rm r-modes} = 1\times 10^{3}$, and $K_{\rm magnetars} = 1\times 10^{-11}$, implying $\Omega_{\rm ref, \, BNS} \simeq 2.12\times 10^{-7}$, $\Omega_{\rm ref, \, r-modes} \simeq 1.69\times 10^{-7}$, and $\Omega_{\rm ref, \, magnetars} \simeq 1.79\times 10^{-8}$, respectively. We have again analysed the data set using the same spectral indices as in the main text ($\alpha = 0,\, 2/3,\, 2,\, 3,\, 4$) to %newly 
evaluate the impact of searching for fewer or more components than the observable ones in this scenario.

Unlike the pure power-law case, we cannot use the $20-1726$ Hz band to analyse this data set. The $\Omega_{\gw, \, i}(f)$ spectra are no longer power-law only in this range, leading to the failure in recovering the injected parameters for every combination of spectral indices. To choose the upper bound in the frequencies to employ in the analysis of this data set, we have made a compromise between the (non-)power-law behaviour of the injected signals in the band and the best recovery of the injected parameters when we inject the signals on top of O3 data. This resulted in the choice of $20-100$ Hz, which we have also used for the analysis in the main text.

\subsubsection{Signal-only case using O3 sensitivity}
Similarly to the previously presented injection studies, we have considered a dataset with only the signal injected. In this specific case, the dataset serves two scopes: first, testing the multi-components analysis when multiple SGWBs are present in the data and they differ in intensity, but there is not a dominant one; and second, assessing the generalisation of this method to the inference of the ensemble properties of the GW sources generating the SGWBs.

We see from the joint-estimation results in table \ref{tab:astro_injections_estimator_table_no_noise} that one retrieves the intensities of the injected $\Omega_{\alpha}$ are recovered again when one considers the combination $\alpha = 2/3, \, 2, \, 4$. 
This also means that the power-law approximation of the signals is valid in the considered frequency band.
However, it is worth observing that, in contrast to the previous signal-only datasets, the estimator is not perfectly centred on the injected values, as confirmed by the PE results in figure \ref{fig:BNS_injection_PE_plots}. 
The effects of the non-power-law injections are also visible for other spectral indices combinations, making acquiring non-zero values for the zero-components $\alpha = 0$ and $\alpha = 3$ even when they are considered together with $\alpha = 2/3, \, 2, \, 4$. 

\begin{table*}
\scriptsize
\makebox[\textwidth][c]{
\begin{tabular}{llllll}
\toprule
{} &               $\hat{\Omega}_{0}$ &            $\hat{\Omega}_{2/3}= 2.1\times 10^{-7}$ &              $\hat{\Omega}_{2}= 1.7\times 10^{-7}$ &              $\hat{\Omega}_{3}$ &              $\hat{\Omega}_{4}= 1.8\times 10^{-8}$ \\
\midrule
$\alpha=\{0\}$               &        $(8.66 \pm0.08)\times 10^{-7}$ &                               - &                               - &                                    - &                               - \\
$\alpha=\{2/3\}$             &                                     - &  $(7.11 \pm0.06)\times 10^{-7}$ &                               - &                                    - &                               - \\
$\alpha=\{2\}$               &                                     - &                               - &  $(3.62 \pm0.03)\times 10^{-7}$ &                                    - &                               - \\
$\alpha=\{3\}$               &                                     - &                               - &                               - &       $(1.45 \pm0.01)\times 10^{-7}$ &                               - \\
$\alpha=\{4\}$               &                                     - &                               - &                               - &                                    - &  $(4.32 \pm0.03)\times 10^{-8}$ \\
\hline
$\alpha=\{0, 2/3\}$          &       $(-2.72 \pm0.05)\times 10^{-6}$ &  $(2.71 \pm0.03)\times 10^{-6}$ &                               - &                                    - &                               - \\
$\alpha=\{0, 2\}$            &         $(-1.4 \pm0.1)\times 10^{-7}$ &                               - &  $(4.02 \pm0.05)\times 10^{-7}$ &                                    - &                               - \\
$\alpha=\{0, 3\}$            &         $(3.5 \pm0.1)\times 10^{-7}$ &                               - &                               - &       $(1.16 \pm0.01)\times 10^{-7}$ &                               - \\
$\alpha=\{0, 4\}$            &        $(5.73 \pm0.08)\times 10^{-7}$ &                               - &                               - &                                    - &  $(3.26 \pm0.04)\times 10^{-8}$ \\
$\alpha=\{2/3, 2\}$          &                                     - &   $(-1.6 \pm0.1)\times 10^{-7}$ &  $(4.29 \pm0.06)\times 10^{-7}$ &                                    - &                               - \\
$\alpha=\{2/3, 3\}$          &                                     - &  $(2.94 \pm0.08)\times 10^{-7}$ &                               - &       $(1.06 \pm0.01)\times 10^{-7}$ &                               - \\
$\alpha=\{2/3, 4\}$          &                                     - &  $(4.67 \pm0.07)\times 10^{-7}$ &                               - &                                    - &  $(2.84 \pm0.04)\times 10^{-8}$ \\
$\alpha=\{2, 3\}$            &                                     - &                               - &  $(2.53 \pm0.07)\times 10^{-7}$ &         $(4.8 \pm0.3)\times 10^{-8}$ &                               - \\
$\alpha=\{2, 4\}$            &                                     - &                               - &  $(3.03 \pm0.04)\times 10^{-7}$ &                                    - &    $(9.9 \pm0.6)\times 10^{-9}$ \\
$\alpha=\{3, 4\}$            &                                     - &                               - &                               - &        $(2.60 \pm0.04)\times 10^{-7}$ &   $(-4.0 \pm0.1)\times 10^{-8}$ \\
\hline
$\alpha=\{0, 2/3, 2\}$       &          $(1.5 \pm0.1)\times 10^{-6}$ &   $(-1.6 \pm0.1)\times 10^{-6}$ &    $(6.1 \pm0.2)\times 10^{-7}$ &                                    - &                               - \\
$\alpha=\{0, 2/3, 3\}$       &          $(7.6 \pm8.3)\times 10^{-8}$ &    $(2.3 \pm0.7)\times 10^{-7}$ &                               - &       $(1.08 \pm0.03)\times 10^{-7}$ &                               - \\
$\alpha=\{0, 2/3, 4\}$       &         $(-5.9 \pm0.7)\times 10^{-7}$ &    $(9.3 \pm0.6)\times 10^{-7}$ &                               - &                                    - &  $(2.46 \pm0.06)\times 10^{-8}$ \\
$\alpha=\{0, 2, 3\}$         &          $(2.6 \pm0.3)\times 10^{-7}$ &                               - &    $(6.7 \pm2.3)\times 10^{-8}$ &         $(9.7 \pm0.6)\times 10^{-8}$ &                               - \\
$\alpha=\{0, 2, 4\}$         &          $(1.8 \pm0.2)\times 10^{-7}$ &                               - &    $(2.2 \pm0.1)\times 10^{-7}$ &                                    - &    $(1.6 \pm0.1)\times 10^{-8}$ \\
$\alpha=\{0, 3, 4\}$         &          $(3.1 \pm0.2)\times 10^{-7}$ &                               - &                               - &       $(1.35 \pm0.08)\times 10^{-7}$ &   $(-5.6 \pm2.3)\times 10^{-9}$ \\
$\alpha=\{2/3, 2, 3\}$       &                                     - &    $(3.4 \pm0.4)\times 10^{-7}$ &   $(-4.4 \pm3.4)\times 10^{-8}$ &       $(1.17 \pm0.08)\times 10^{-7}$ &                               - \\
$\alpha=\{2/3, 2, 4\}$       &                                     - &    $(2.1 \pm0.3)\times 10^{-7}$ &    $(1.7 \pm0.2)\times 10^{-7}$ &                                    - &    $(1.8 \pm0.1)\times 10^{-8}$ \\
$\alpha=\{2/3, 3, 4\}$       &                                     - &    $(3.2 \pm0.2)\times 10^{-7}$ &                               - &         $(9.0 \pm1.0)\times 10^{-8}$ &    $(4.5 \pm2.8)\times 10^{-9}$ \\
$\alpha=\{2, 3, 4\}$         &                                     - &                               - &    $(4.9 \pm0.3)\times 10^{-7}$ &        $(-1.6 \pm0.2)\times 10^{-7}$ &    $(4.2 \pm0.5)\times 10^{-8}$ \\
\hline
$\alpha=\{0, 2/3, 2, 3\}$    &         $(-4.8 \pm3.5)\times 10^{-7}$ &    $(9.7 \pm4.6)\times 10^{-7}$ &   $(-2.4 \pm1.5)\times 10^{-7}$ &         $(1.5 \pm0.3)\times 10^{-7}$ &                               - \\
$\alpha=\{0, 2/3, 2, 4\}$    &   $(0.004 \pm2.735)\times 10^{-7}$ &    $(2.1 \pm3.3)\times 10^{-7}$ &    $(1.7 \pm0.8)\times 10^{-7}$ &                                    - &    $(1.8 \pm0.3)\times 10^{-8}$ \\
$\alpha=\{0, 2/3, 3, 4\}$    &         $(-2.1 \pm1.8)\times 10^{-7}$ &    $(5.3 \pm1.9)\times 10^{-7}$ &                               - &         $(6.1 \pm2.8)\times 10^{-8}$ &    $(1.1 \pm0.6)\times 10^{-8}$ \\
$\alpha=\{0, 2, 3, 4\}$      &          $(1.4 \pm0.6)\times 10^{-7}$ &                               - &    $(2.8 \pm1.0)\times 10^{-7}$ &        $(-3.9 \pm6.2)\times 10^{-8}$ &    $(2.2 \pm1.0)\times 10^{-8}$ \\
$\alpha=\{2/3, 2, 3, 4\}$    &                                     - &    $(2.1 \pm1.0)\times 10^{-7}$ &    $(1.7 \pm1.5)\times 10^{-7}$ &    $(0.01 \pm7.84)\times 10^{-8}$ &    $(1.8 \pm1.2)\times 10^{-8}$ \\
\hline
$\alpha=\{0, 2/3, 2, 3, 4\}$ &  $(-0.008 \pm8.945)\times 10^{-7}$ &   $(2.1 \pm13.6)\times 10^{-7}$ &    $(1.7 \pm7.1)\times 10^{-7}$ &  $(0.004 \pm 2.563)\times 10^{-7}$ &    $(1.8 \pm3.0)\times 10^{-8}$ \\
\bottomrule
\end{tabular}
}
\caption{Estimators from the multi-components analysis in the $20-100$ Hz band for the different combinations of the five spectral indices for the astrophysical BNS, r-mode, and magnetar SGWB signal-only injection data set, with injected parameters $K_{\rm CBC}= 7.91 \times 10^{5}\, M_{\odot}\, {\rm Gpc^{-3}\, yr^{-1}}$, $K_{\rm r-modes}=1\times 10^{3}$, and $K_{\rm magnetars}= 1\times 10^{-11}\, {\rm T^{-1}}$. Horizontal lines divide the table in regions where a fixed number of components is considered for the analysis.}
\label{tab:astro_injections_estimator_table_no_noise}
\end{table*}

Finally, when reinterpreting the analysis results in terms of the injected ensemble properties for $\alpha = 2/3, \, 2, \, 4$, up to small deviations due to residual non-power-law deviations, as visible in figure \ref{fig:BNS_injection_PE_plots}. This shows the robustness of the multi-component approach and of the assumptions made here.

\subsubsection{On top of O3 data}
\begin{figure*}
    \centering
    \includegraphics[width = 0.49\textwidth]{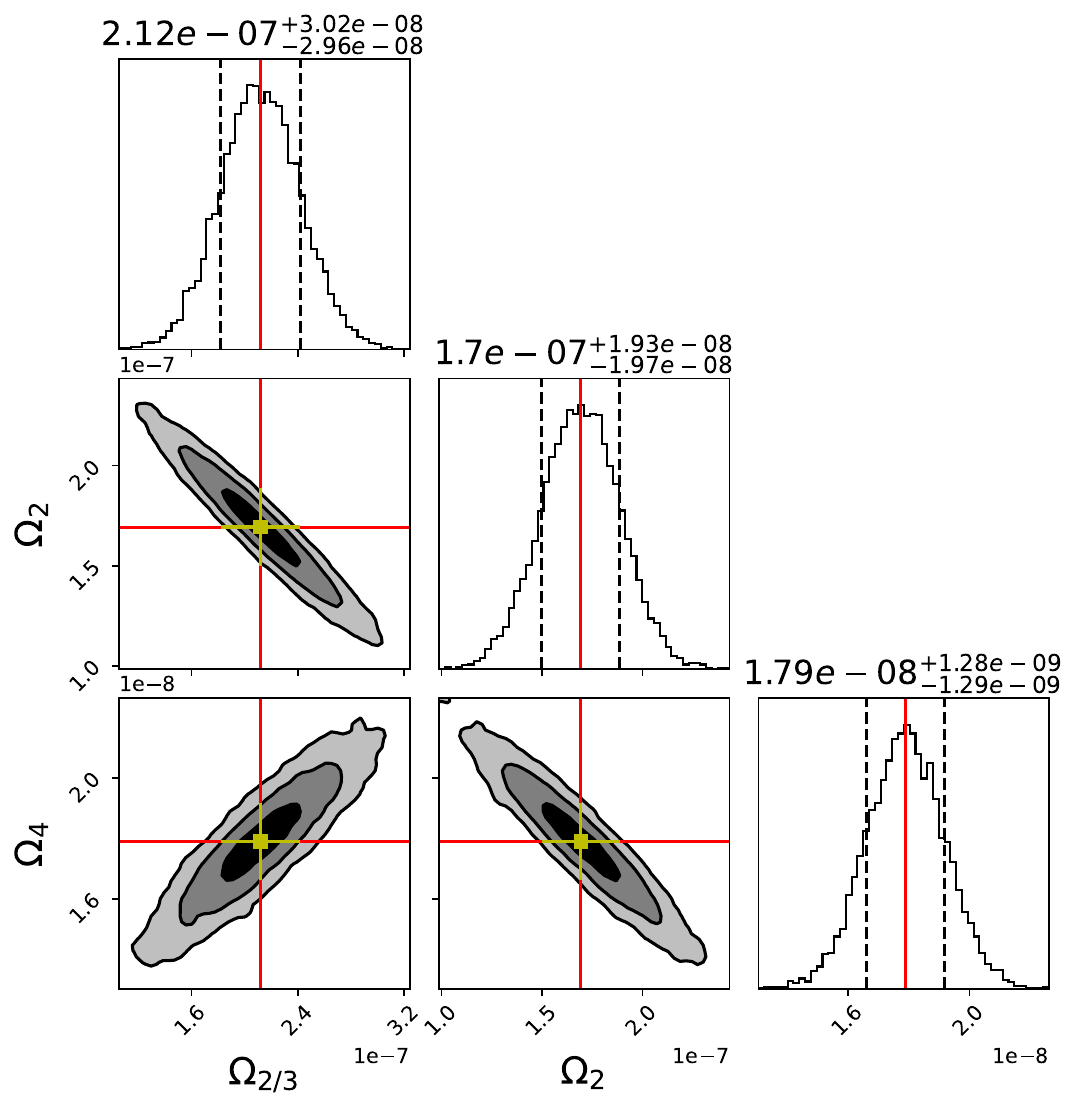}
    \includegraphics[width = 0.49\textwidth]{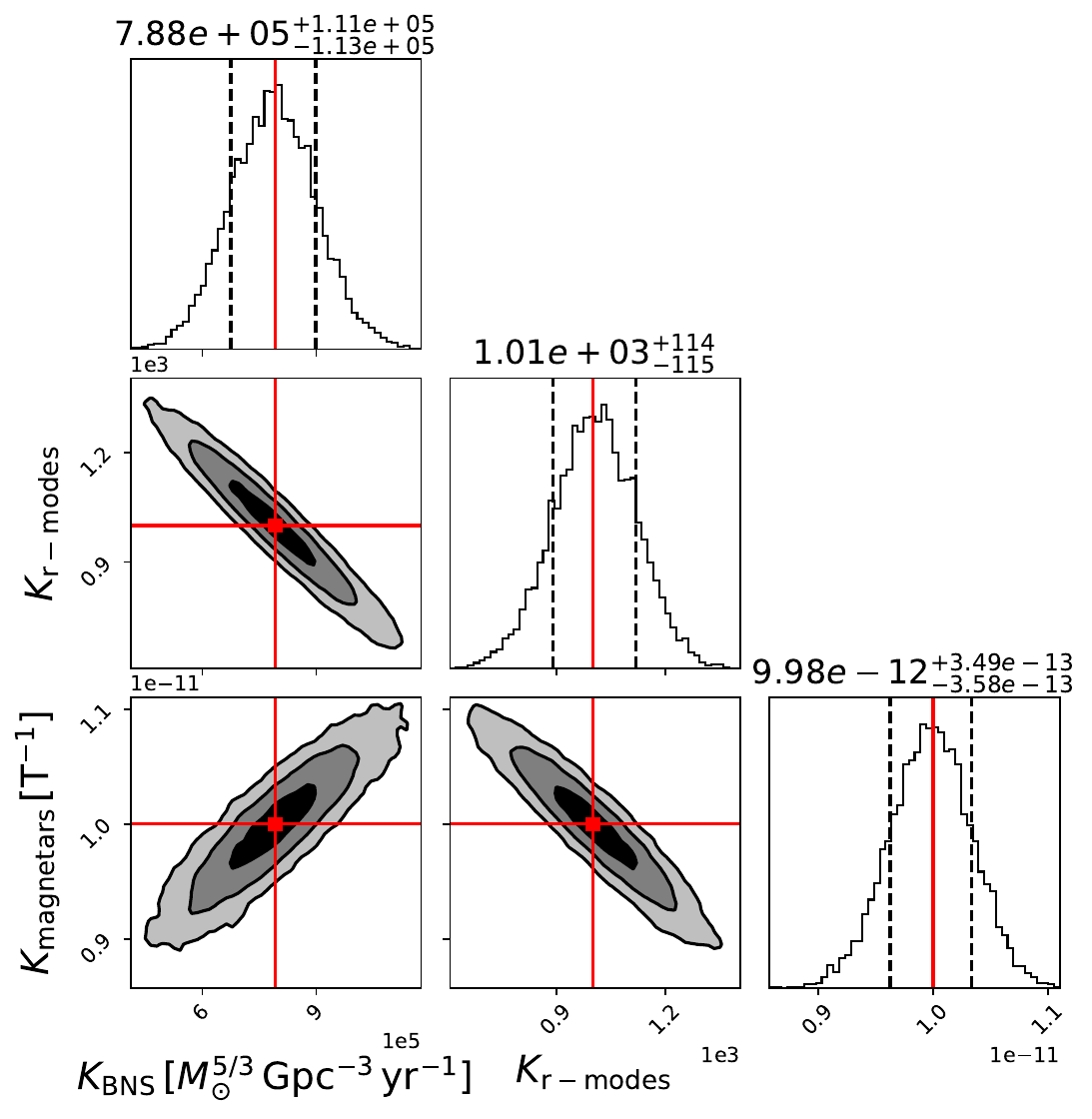}
    \includegraphics[width = 0.49\textwidth]{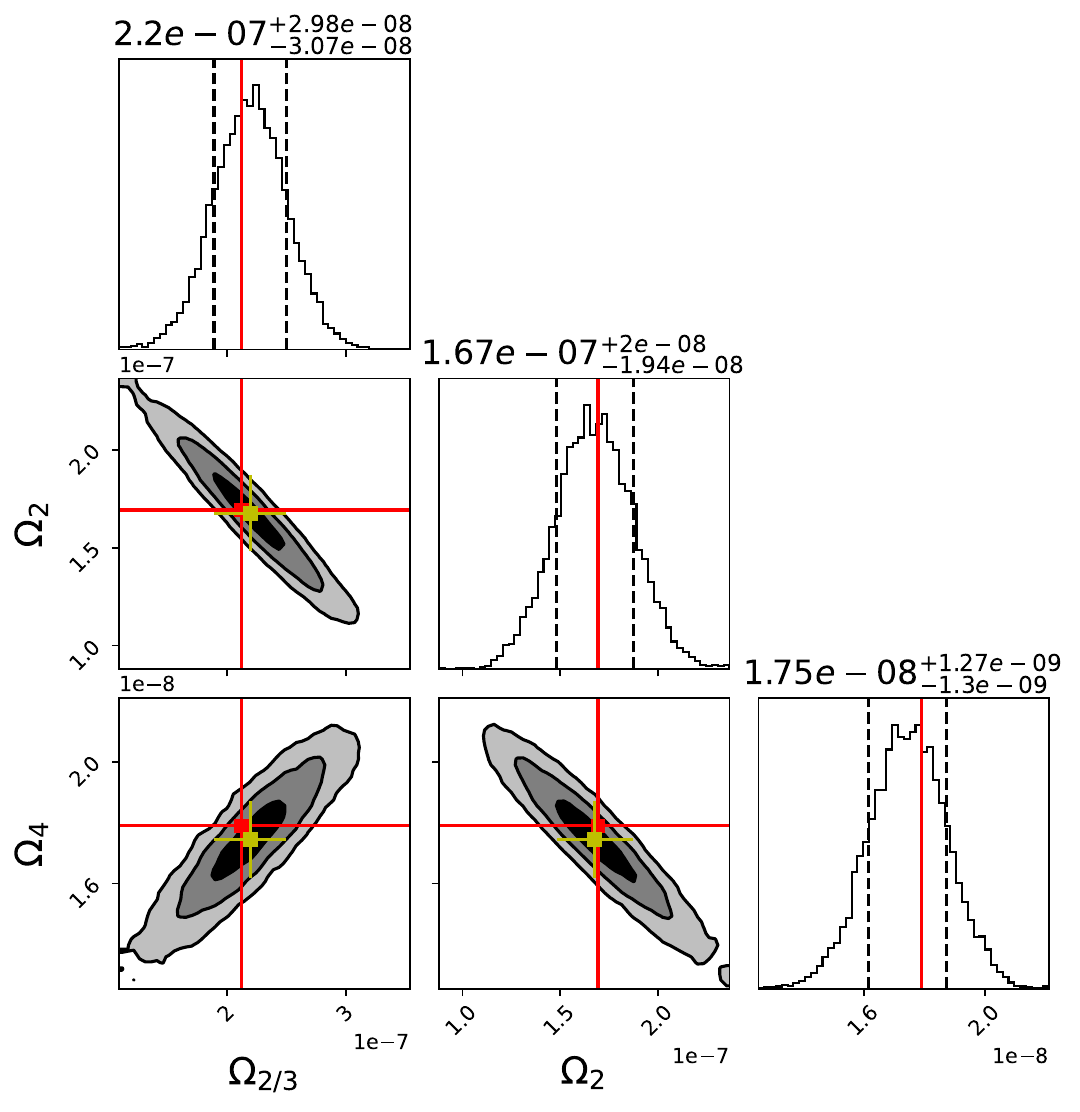}
    \includegraphics[width = 0.49\textwidth]{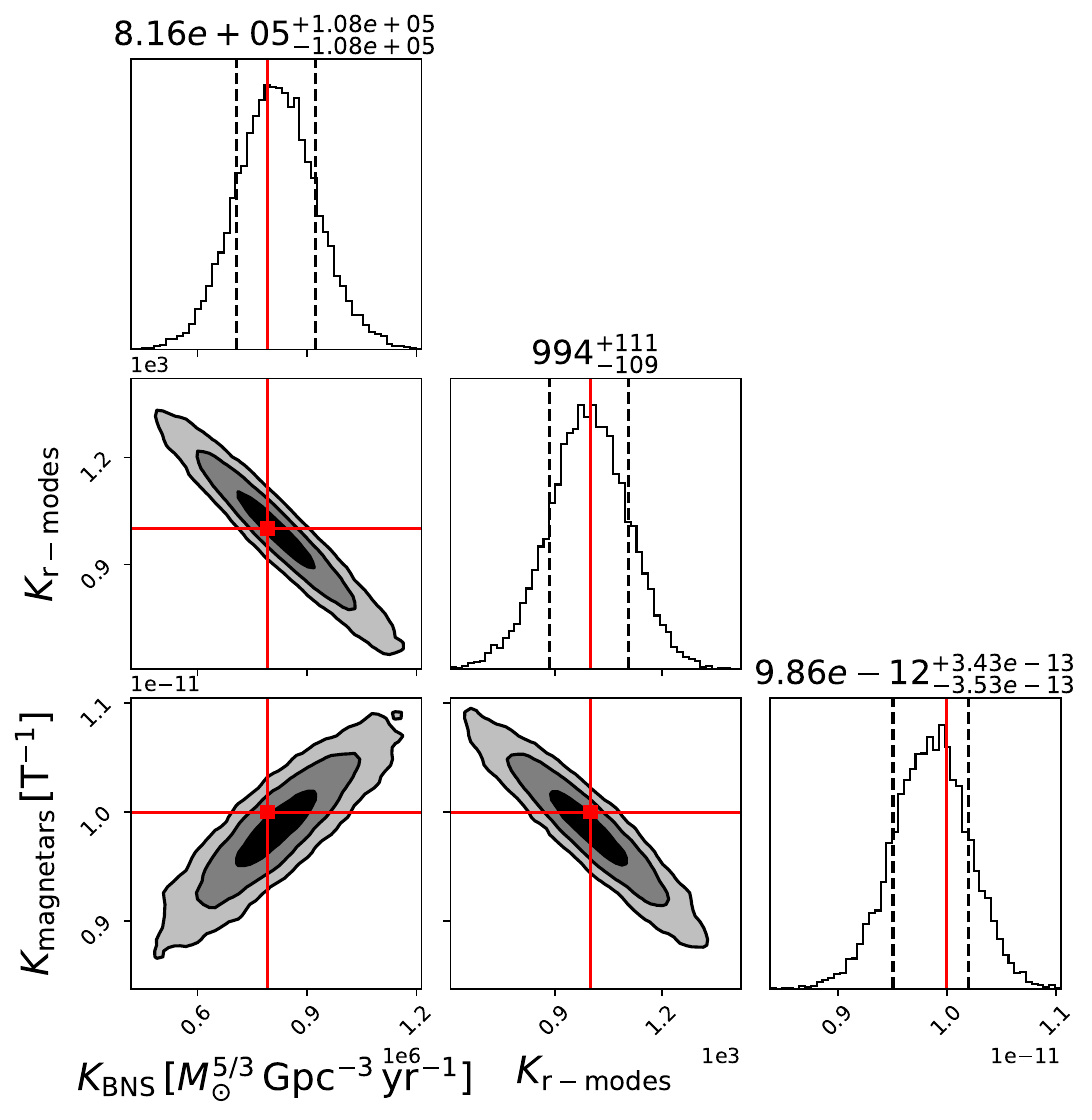}
    \caption{Results of the parameter estimation for the $\alpha = 2/3,\, 2,\,4$ combination in the $20-100$ Hz band for the astrophysical injection for BNS, r-modes, and magnetars SGWBs (top row: signal-only injection; bottom row: injection on top of O3 data). The left panel shows the recovery of the $\Omega_{\alpha}$, while the right panel the injected ensemble parameters. Contour plots show the $1\sigma$, $2\sigma$, and $3\sigma$ credible areas (black, grey, light grey, respectively). The red lines denote the injected values, while the yellow error bars represent the $1\sigma$ uncertainty of the $\Omega_{\alpha}$ estimators from the joint analysis. The dashed black lines in the histogram panels delimit the $1\sigma$ region of the estimated parameters.}
    \label{fig:BNS_injection_PE_plots}
\end{figure*}

Finally, we inject once more the signals on top of O3 data.% and summarise the estimation results in table \ref{tab:BNS_injection_Om_estimator_table_20_100_Hz}. 
In agreement with what was learned from the study of the previous datasets with the injections on top of O3 data, the recovery of the signal and of the injected ensemble parameters happens only when performing the multi-component analysis for the set of spectral indices $\alpha = 2/3,\, 2,\, 4$, as illustrated in figure \ref{fig:BNS_injection_PE_plots}. Once more, the effect of the unknown component $\Omega_{O3}$ introduces bias in the recovery. However, the bias in this dataset is less visible than in the previous cases in sections \ref{sec:equal_PL_injections} and \ref{sec:dominant_PL_injection}. 

\subsection{Take-away message from the injection study}
The injection study has shown that in spite of the biases when missing some components, the multi-component analysis is most robust compared to the single-component analysis, minimising the bias when it is present. The single-component analysis may be the fastest and easiest to interpret in the case where one has the certainty that one of the components is dominant above all the others by multiple orders of magnitude. In such a case, the multi-component analysis can still be used to refine the results of the single-component analysis for the dominant component, helping to identify possible noise sources that contribute to the single-component estimator bias. In addition to that, if the dominant component cannot be correctly subtracted from the data, the multi-component analysis proves to be fundamental to correctly identifying the subdominant components, whose estimate would otherwise be noticeably biased. 

\end{document}